\documentclass[11pt,a4paper,twocolumn]{article}
\usepackage[utf8]{inputenc}
\usepackage{amsmath}
\usepackage{amsfonts}
\usepackage{amssymb}
\usepackage{subfig}
\usepackage{graphicx}
\usepackage[labelfont=bf]{caption}
\usepackage[left=1.5cm,right=1.5cm,top=2cm,bottom=2cm]{geometry}
\usepackage[style=nature,natbib=true,backend=biber, doi=false,isbn=false,url=false,date=year,maxbibnames=20]{biblatex}

\DeclareFieldFormat*{title}{#1}

\usepackage[switch]{lineno}

\usepackage[colorlinks=false, linkcolor=blue]{hyperref}
\usepackage[small]{titlesec}
\usepackage{dblfloatfix}
\usepackage[table]{xcolor}
\usepackage{microtype}
\usepackage{sidecap}
\usepackage{makecell}

\definecolor{tblcolor}{HTML}{f0f6f7}

\title{\vspace{-0.5em}What is a meaningful representation of protein sequences?}

\author{
    \small Nicki Skafte Detlefsen \\
    \small Section for Cognitive Systems \\
    \small Technical University of Denmark \\
    \small \texttt{nsde@dtu.dk} \\
    \and
    \small Søren Hauberg \\
    \small Section for Cognitive Systems \\
    \small Technical University of Denmark \\
    \small \texttt{sohau@dtu.dk}
    \and
    \small Wouter Boomsma \\
    \small Department of Computer Science \\
    \small University of Copenhagen \\
    \small \texttt{wb@di.ku.dk} \\
}

\date{} 

\newcommand{\norm}[1]{\left\lVert#1\right\rVert}
\newcommand{\fig}{Figure~}
\newcommand{\tab}{Table~}
\newcommand{\X}{\textbf{X}}
\newcommand{\Z}{\textbf{Z}}

\titlespacing*\subsection{0pt}{12pt plus 4pt minus 2pt}{4pt plus 2pt minus 2pt}

\begin{document}
\maketitle

\begin{abstract}
How we choose to represent our data has a fundamental impact on our ability to subsequently extract information from them. Machine learning promises to automatically determine efficient representations from large unstructured datasets, such as those arising in biology. However, empirical evidence suggests that seemingly minor changes to these machine learning models yield drastically different data representations that result in different biological interpretations of data. This begs the question of what even constitutes the most meaningful representation. Here, we approach this question for representations of protein sequences, which have received considerable attention in the recent literature.
We explore two key contexts in which representations naturally arise: transfer learning and interpretable learning. In the first context, we demonstrate that several contemporary practices yield suboptimal performance, and in the latter we demonstrate that taking representation geometry into account significantly improves interpretability and lets the models reveal biological information that is otherwise obscured.
\end{abstract}
\vspace{-0.4cm}
\section*{Introduction}

\emph{Data representations} play a crucial role in the statistical analysis of biological data.
At its core, a representation\footnote{We note that `representation', `feature' and `embedding' all seem to be used interchangeably in the literature.} is a distillation of raw data into an abstract, high-level and often lower dimensional space that captures the essential features of the original data. This can subsequently be used for data exploration, e.g.\ through visualization, or task-specific predictions where limited data is available. Given the importance of representations
it is no surprise that we see 
a rise in biology of \emph{representation learning} \cite{Bengio2013}, a subfield of machine learning where the representation is estimated alongside the statistical model.
In the analysis of protein sequences in particular, the last years have produced a number of studies that demonstrate how representations can help extract important biological information automatically from the millions of observations acquired through modern sequencing technologies \cite{Riesselman2018, bepler2019learning, Alley2019, Rao2019, rives2019biological, shin2021protein, Heinzinger2019, madani2020progen, elnaggar2020prottrans,Lu2020.09.04.283929,frazer2020large,10.1093/bioinformatics/btab083,repecka2021expanding}. While these promising results indicate that learned representations can have substantial impact on scientific data analysis, they also beg the question: \emph{what is a good representation?} This elementary question is the focus of this paper. \looseness=-1

A classic example of representation learning is 
principal component analysis (PCA) \cite{Jolliffe:1986}, which learns features that are linearly related to the original data. 
Contemporary techniques dispel with the assumption of linearity and instead seek highly non-linear relations \cite{Bengio2013}, often by employing neural networks. This has been particularly successful in natural language processing (NLP), where representations of word sequences are learned from vast online textual resources, extracting general properties of language that support subsequent specific language tasks \cite{Radford2018, Devlin2019, Liu2019}. The success of such \emph{word sequence models} has inspired its use for modelling \emph{biological sequences}, leading to impressive results in application areas such as remote homologue detection \cite{morton2020protein}, function classification \cite{gligorijevic2020structure} and prediction of mutational effects \cite{rives2019biological}. 

\begin{figure*}
\centering
\includegraphics[width=\textwidth]{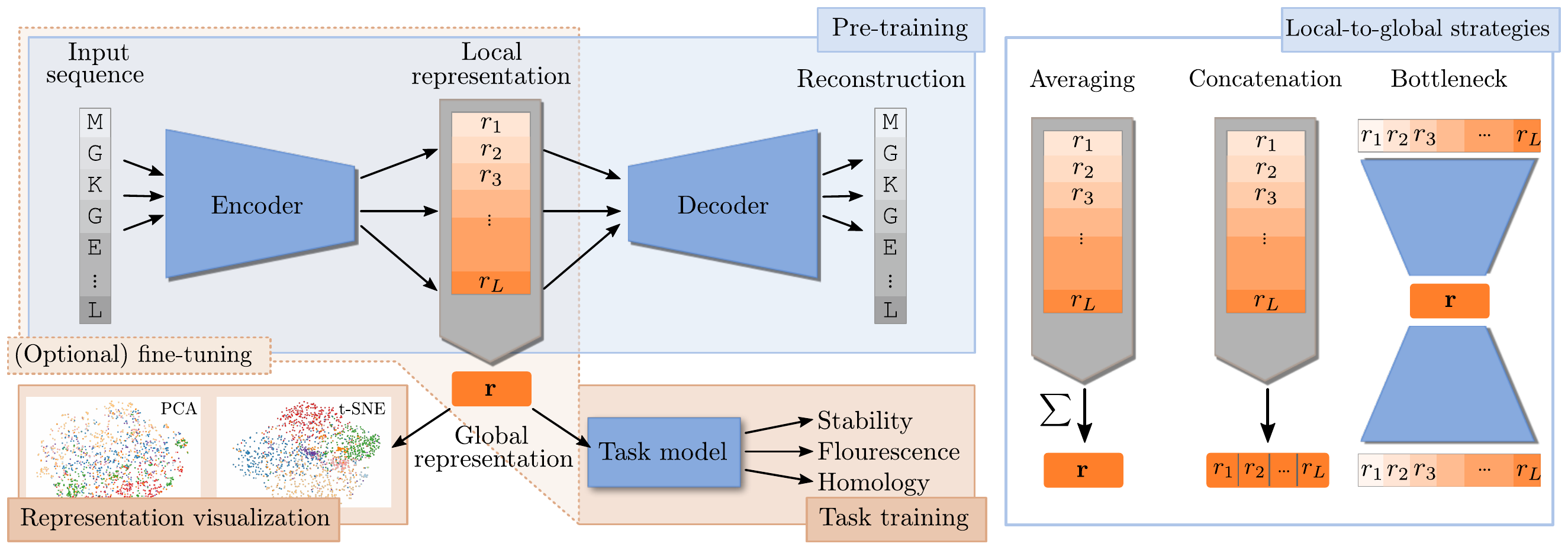}
\caption{
Representations of protein sequences. 
During the \textit{pre-training} phase, a model is trained to \textit{embed} or \emph{encode} input protein sequences $(s_1,s_2,...,s_L)$, to a local representation $(r_1,r_2,...,r_L)$, after which
it is \emph{decoded} to be as similar as possible to the original sequence. After the pre-training stage, the learned representation can be used as a proxy for the raw input sequence, either for direct visual interpretation, or 
as input to a supervised model trained for a specific task (transfer-learning). When working in the transfer-learning setting, it is possible to also update the parameters of the encoder while training on the specific task, thereby fine-tuning the representation to the task of interest.
For interpretation or for prediction of global properties of proteins, the local representations $r_i$, are aggregated into a global representation, often using a simple procedure such as averaging over the sequence length. For visualization purposes these global representations are then often dimensionality reduced using standard procedures such as PCA or t-SNE. }

\label{fig:embed_task_overview}
\end{figure*}

Since representations are becoming an important part of biological sequence analysis, we should think critically about whether the constructed representations efficiently capture the information we desire. This paper discusses this topic, with focus on protein sequences, although many of the insights apply to other biological sequences as well \cite{10.1093/bioinformatics/btab083}. 
Our work consists of two parts. First, we consider representations in the transfer-learning setting. We investigate the impact of network design and training protocol on the resulting representation, and find that several current practices are suboptimal. Second, we investigate the use of representations for the purpose of data interpretation. We show that explicit modeling of the \emph{representation geometry} allows us to extract robust and identifiable biological conclusions. Our results demonstrate a clear potential for \emph{designing} representations actively, and for \emph{analyzing} them appropriately.

\section*{Results}
Representation learning has at least two uses: In \emph{transfer learning} we seek a representation that improves a downstream task, and in \emph{data interpretation} the representation should reveal the data's underlying patterns, e.g.\ through visualization. Since the first has been at the center of recent literature \cite{Alley2019, Armenteros2020,Heinzinger2019,Rao2019,madani2020progen, elnaggar2020prottrans,gligorijevic2020structure}, we place our initial focus there, and turn later to data interpretation.

\subsection*{Representations for transfer learning}
Transfer learning addresses the problems caused by limited access to labeled data. For instance,  when predicting the stability of a given protein, we only have limited training data available as it is experimentally
costly to measure stability. The key idea is to leverage the many available unlabeled protein sequences
to learn (\emph{pre-train}) a general protein representation through an \emph{embedding model}, and then train a problem-specific
\emph{task model} on top using the limited labeled training data (\fig\ref{fig:embed_task_overview}). 

In the protein setting, learning representations for transfer-learning can be implemented at different scopes. It can be addressed at a universal scope, where representations are learned to reflect general properties of all proteins, or it can be implemented at the scope of an individual protein family, where an embedding model is pre-trained only on closely related sequences. Initially, we will focus on universal setting, but will return to family specific models in the second half of the paper.

\begin{table*}[t!]
\centering
\rowcolors{2}{white}{tblcolor}
\begin{tabular}{l|ccc|ccc|ccc}
        & \multicolumn{3}{c|}{Remote Homology} & \multicolumn{3}{c|}{Fluorescence} & \multicolumn{3}{c}{Stability} \\ \cline{2-10} 
        & Resnet    & LSTM    & Trans    & Resnet   & LSTM   & Trans   & Resnet  & LSTM  & Trans \\ \hline
\textsc{Pre\hspace{0.5mm}+Fix} & 0.27          & \textbf{0.37}        & 0.27               & 0.23         & \textbf{0.74}       & 0.48              & 0.65        & 0.70      & 0.62            \\
\textsc{Pre\hspace{0.5mm}+Fin} & 0.17          & 0.26        & 0.21               & 0.21         &  0.67      & 0.68              & \textbf{0.73}        & 0.69      & \textbf{0.73}  \\
\textsc{Rng+Fix} & 0.03          & 0.10        & 0.04               & 0.25         & 0.63       & 0.14              & 0.21        & 0.61       & -            \\
\textsc{Rng+Fin} & 0.10          & 0.12        & 0.09               & -0.28         & 0.21       & 0.22              & 0.61         & 0.28      & -0.06            \\
\hline
Baseline & \multicolumn{3}{c|}{0.09 (Accuracy)}                 & \multicolumn{3}{c|}{0.14 (Correlation)}             & \multicolumn{3}{c}{0.19 (Correlation)}         
\end{tabular}
\caption{The impact of fine-tuning and initialization on downstream model performance.
The embedding models were either randomly initialized (\textsc{Rng}) or pre-trained (\textsc{Pre}), and subsequently either fixed (\textsc{Fix}) or fine-tuned to the task (\textsc{Fin}). The baseline is a simple one-hot encoding of the sequence. Although fine-tuning is beneficial on some task/model combinations, we see clear signs of overfitting in the majority of cases (best results in bold).}
\label{tab:downstream_results}
\end{table*}

When considering representations in the transfer-learning setting, the quality, or \emph{meaningfulness}, of a representation is judged merely by the level of predictive performance obtained by one or more downstream tasks. Our initial task will therefore be to study how this performance depends on common modeling assumptions.
A recent study established a benchmark set of predictive tasks for protein sequence representations \cite{Rao2019}. For our experiments below, we will consider three of these tasks, each reflecting a particular global protein property: 1) classification of protein sequences into a set of 1,195 known folds \cite{Hou2018}, 2) fluorescence prediction for variants of the green fluorescent protein in \emph{Aequorea victoria} \cite{Sarkisyan2016}, and 3) prediction of the stability of protein variants obtained in high throughput experimental design experiments \cite{Rocklin2017}. 

\paragraph{Fine-tuning can be detrimental to performance. }
In the transfer-learning setting, 
the pre-training phase and the task learning phase are conceptually separate (\fig\ref{fig:embed_task_overview}, left), but 
it is common practice to fine-tune the embedding model for a given task, which implies that the parameters of both models are in fact optimized jointly \cite{Rao2019}.  
Given the large number of parameters typically employed in embedding models, we hypothesize that this can lead to overfitted representations, 
at least in the common scenario where only limited data is available for the task learning phase.

To test this hypothesis we train three models, an LSTM \cite{Hochreiter1997}, a Transformer \cite{Vaswani2017}, and a dilated residual network (Resnet) \cite{Yu2017} on a diverse set of protein sequences extracted from Pfam\cite{pfam}, where we either keep the embedding model fixed (\textsc{Fix}) or fine-tune it to the task (\textsc{Fin}). To evaluate the impact of the representation model itself, we consider both a pre-trained version (\textsc{Pre}) and randomly initialized representation models that are not trained on data (\textsc{Rng}). Such models will map similar inputs to similar representations, but should otherwise not perform well.
Finally, as a naive baseline representation, we consider the direct one-hot encoding of each amino acid in the sequence. In all cases, we extract global representations using an attention-based averaging over local representations (\fig\ref{fig:embed_task_overview}, right).

Table~\ref{tab:downstream_results} shows that fine-tuning the embedding clearly reduces test performance in two out of three tasks, confirming that fine-tuning can have significant detrimental effects in practice. Incidentally, we also note that the randomly initialized representation performs remarkably well in several cases, which echoes results known from random projections \cite{Bingham2001}.

\emph{Implication: fine-tuning a representation to a specific task carries the risk of overfitting, since it often increases the number of free parameters substantially, and should therefore take place only under rigorous cross validation. Fixing the embedding model during task-training should be the default choice.}

\paragraph{Constructing a global representation as an average of local representations is suboptimal.}

One of the key modeling choices for biological sequences is how to handle their sequential nature. Inspired by developments in natural language processing, most of the recent representation learning advances for proteins use language models, which aim to reproduce their own input, either by predicting the next character given the sequence observed so far, or by predicting the entire sequence from a partially obscured input sequence. 
The representation learned by such models is a sequence of \emph{local representations} $(r_1, r_2, ..., r_L)$ each corresponding to one amino acid in the input sequence $(s_1, s_2, ..., s_L)$. To successfully predict the next amino acid, $r_i$ should contain information about the local neighborhood around $s_i$, together with some global signal reflecting properties of the complete sequence.
In order to obtain a global representation of the entire protein, the variable number of local representations must be aggregated into a fixed-size global representation. \emph{A priori}, we would expect this choice to be quite critical to the nature of the resulting representation. Standard approaches for this operation include averaging with uniform \cite{Alley2019,elnaggar2020prottrans} or learned attention \cite{Rao2019, light_attention, monteiro-etal-2020-performance} weights or simply using the maximum value. However, the complex non-local interactions known to occur in a protein suggest that it could be beneficial to allow for more complex aggregation functions. To investigate this issue, we consider two alternative strategies (\fig\ref{fig:embed_task_overview}, right):

The first strategy (Concat) avoids aggregation altogether by \emph{concatenating} the local representations $r=[r_1, r_2, ..., r_L, p, p, p]$ (with additional padding $p$ to adjust for variable sequence-length). This approach preserves all information stored in the local $r_i$s. To make a fair comparison to the averaging strategy, we maintain the same overall representation size by scaling down the size of of the local representations $r_i$. In our case, with a global representation size of 2048, and a maximal sequence length of 512, this means that we restrict the local representation to only four dimensions. 

As a second strategy (Bottleneck), we investigate the possibility of \emph{learning} the optimal aggregation operation, using an autoencoder, a simple neural network that as output predicts its own input, but forces it through a low-dimensional bottleneck  \cite{Kramer1991}. The model thus \emph{learns} a generic global representation during pre-training, in contrast to the strategies above in which the global representation arises as a deterministic operation on the learned local representations. We implement the Bottleneck strategy within the Resnet (convolutional) setting, where we have well-defined procedures for down- and upsampling the sequence length.

When comparing the two proposed aggregation strategies on the three protein prediction tasks (Stability, Fluorescence, Remote Homology), we observe a quite dramatic impact on performance (\tab\ref{tab:sequence_length_results}). The Bottleneck strategy, where the global representation is learned, clearly outperforms the other strategies. This was expected, since already during pre-training this model is encouraged to find a more global structure in the representations. More surprising are the results for the Concat strategy, as these demonstrate that even if we restrict the local representation to be much smaller than in standard sequential models, the fact that there is no loss of information during aggregation has a significant positive influence on the downstream performance. 

\emph{Implication: if a global representation of proteins is required, it should be learned rather than calculated as an average of local representations.}

\begin{table}[h!]
    \rowcolors{2}{white}{tblcolor}
    \begin{tabular}{l|ccc}
	  	                & \makecell{Stability \\ (Corr.)} & \makecell{Fluorescence \\ (Corr.)}  & \makecell{Homology \\ (Acc.)} \\ \hline
    Mean            	& 0.42      & 0.19         & 0.27     \\
    Attention           & 0.65      & 0.23         & 0.27     \\
    Light Att.          & 0.66      & 0.23         & 0.27    \\
    Maximum             & 0.02      & 0.02         & 0.28     \\
    MeanMax             & 0.37      & 0.15         & 0.26     \\
    KMax                & 0.10      & 0.11         & 0.27     \\
    Concat     	        & 0.74      & 0.69         & 0.34     \\
    Bottleneck          & \textbf{0.79}      & \textbf{0.78}         & \textbf{0.41}
    \end{tabular}
    \caption{Comparison of strategies for obtaining global, sequence-length independent representations on three downstream tasks \cite{Rao2019}. The first six are variants of averaging used in the literature, using uniform weights (Mean), some variant of learned attention weights (Attention \cite{Rao2019}, Light Attention \cite{light_attention}), or averages of the local representation with the highest attention weight (Maximum, MeanMax, KMax(K=5)). They all use the same pretrained and backbone Resnet model, while the last two
    entries use modified Resnet architectures using either a very low-dimensional feature representation (Concat), or an autoencoder-like structure downsample the representation length. In all cases, training proceeded without fine-tuning. The results demonstrate that simple alternatives such as concatenating smaller local representations (Concat) or changing the model to directly learn a global representation (Bottleneck) can have a substantial impact on performance (best results in bold).}
    \label{tab:sequence_length_results}
\end{table}

\paragraph{Reconstruction error is not a good measure of representation quality.}

Any choice of embedding model will have a number of hyper parameters, such as the number of nodes in the neural network or the dimensionality of the representation itself. How do we choose such parameters? A common strategy is to make these choices based on the reconstruction capabilities of the embedding model, but is it reasonable to expect that this is also the optimal choice from the perspective of the downstream task?

\begin{figure*}[ht!] 
    \centering
    \includegraphics[width=\textwidth]{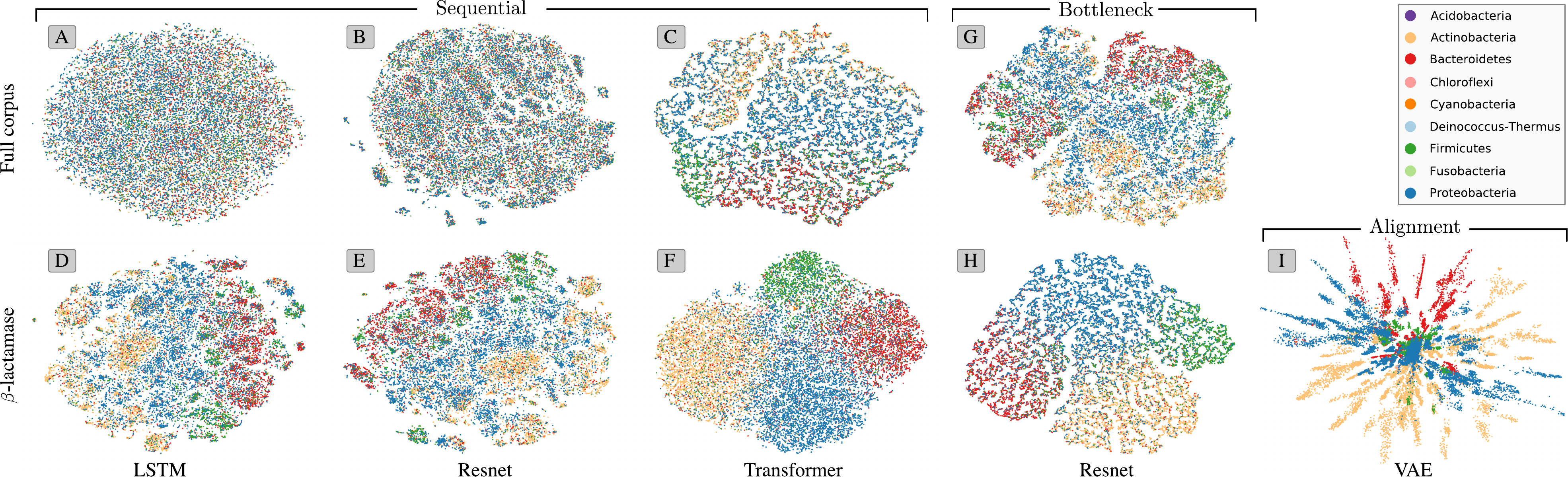}
    \caption{Latent embedding of the protein family of $\beta$-lactamase, color-coded by taxonomy at the phyla level. In the upper row, we embed the family using sequential models (LSTM, Resnet, Transformer) trained on the full corpus of protein families. In the lower row we train the same sequential models again only on the $\beta$-lactamase family (PFAM PF00144 \cite{pfam}). For the models in the first three columns, a simple mean strategy is employed to extract a global representation from local representations, while the fourth column uses the Bottleneck aggregation method. Finally, in the last column, we show the result of preprocessing the sequences in a multiple sequence alignment and applying a dense variational autoencoder (VAE) model. We see clear differences in how well the different phyla are separated, which demonstrates the impact that model choice and data preprocessing can have on the learned representation.}
    \label{fig:beta_lactamase}
\end{figure*}

As an example, we will consider the task of finding the optimal representation size. We trained and evaluated several Bottleneck Resnet models with varying representation dimensions and applied them to the three downstream tasks. 
The results show a clear pattern where the reconstruction accuracy increases monotonically with latent size, with the sharpest increase in the region of 10 to 500, but with marginal improvements all the way up to the maximum size of 10000 (\fig{S1}). However, if we consider the three downstream tasks, we see that the performance in all three cases starts decreasing at around size 500-1000, thus showing a discrepancy between the optimal choice with respect to the reconstruction objective and the downstream task objectives. It is important to stress that the reconstruction accuracy is measured on a validation set, so our observation is not a matter of overfitting to the training data. We employed a carefully constructed 
train/validation/test partition of UniProt \cite{uniprot} provided by Armenteros et al \cite{Armenteros2020}, to avoid overlap between the sets. The results thus show that there is enough data for the embedding model to support a large representation size, while the downstream tasks prefer a smaller input size. The exact behavior will depend on the task and the available data for the two training phases, but we can conclude that there is generally no reason to believe that reconstruction accuracy and the downstream task accuracy will agree on the optimal choice of hyperparameters. Similar findings were reported in the TAPE study \cite{Rao2019}.

\emph{Implication: in transfer-learning, optimal values for hyperparameters (e.g.\ representation size) can in general not be estimated during pre-training. They must be tuned for the specific task.}

\subsection*{Representations for data interpretation}

We now return to the use of representations for data interpretation. If a representation accurately describes structure in the underlying dataset, we might expect it to be useful not only as input to a downstream model, but also as the basis for direct interpretation, for instance through visualization. In this context, it is important to realize that different modeling choices can lead to dramatically different interpretations of the same data. More troubling, even when using the same model assumptions, repeated training instances can also deviate substantially, and we must therefore analyze our interpretations with care. 
In the following, we explore these effects in detail.

\paragraph{Representation manifolds are shaped by scope, model architecture and data preprocessing.}
Recent models for proteins tend to learn universal, cross-family representations of protein space. In bioinformatics, there is, however, a long history of analyzing proteins per family.
Since the proteins in the same family share a common three-dimensional structure, an underlying correspondence exists between positions in different sequences, which we can approximate using multiple sequence alignment techniques. After establishing such an alignment, all input sequences will have the same length, making it possible to use 
simple fixed-size input models, rather than the sequential models discussed previously. One advantage is that models can now readily detect patterns at and correlations between absolute positions of the input, and directly observe both conservation and coevolution. In terms of interpretability, this has clear advantages. An example of this approach is the DeepSequence model \cite{Riesselman2018,frazer2020large}, in which the latent space of a Variational Autoencoder (VAE) was shown to clearly separate the input sequences into different phyla, and capture covariance among sites on par with earlier coevolution methods. We reproduce this result using a VAE on the $\beta$-lactamase family PF00144 from PFAM \cite{pfam}, using a 2 dimensional latent space (\fig\ref{fig:beta_lactamase}I).

If we use the universal, full-corpus, sequence models (LSTM, Resnet, Transformer) and the Bottleneck Resnet from the previous sections to embed the same set of proteins from the $\beta$-lactamase family and use t-SNE \cite{vanDerMaaten2008} to reduce the dimensionality of the protein representations into a two dimensional space, we see no clear phylogenetic separation in the case of LSTM and Resnet, and very little for the Transformer and the Bottleneck Resnet (\fig\ref{fig:beta_lactamase}, top row). The fact that the phyla are much less clearly resolved in these sequential models is perhaps unsurprising, since these models have been trained to represent the space of all proteins, and therefore do not have the same capacity to separate details of a single protein family. Indeed, to compensate for this, recent work has introduced the concept of \emph{evo-tuning}, where a universal representation is fine-tuned on a single protein family \cite{Alley2019,biswas2021low}. 

When \emph{training} exclusively on $\beta$-lactamase sequences (\fig\ref{fig:beta_lactamase}, bottom row) we observe more structure for all models, but only the Transformer and Bottleneck Resnet are able to fully separate the different phyla. Comparing this to an alignment-based VAE model, we still see large differences in protein representations, despite the fact that all models now are trained on the same corpus of proteins.

The observed differences between representations is a combined effect arising from the following factors: 1) the inductive biases underlying the different model architectures, 2) the domain-specific knowledge inserted through preprocessing sequences when constructing an alignment, and 3) the post-processing of representation space to make it amenable to visualization in 2D (\fig\ref{fig:beta_lactamase}A-H were processed using t-SNE, see \fig{S3-S4} for equivalent plots using PCA). Often, these contributions are interdependent, and therefore difficult to disentangle. For instance, the VAE can use a simple model architecture only because the sequences have been preprocessed into an alignment. Likewise, the simplicity of the VAE makes it possible to limit the size of the bottleneck to only 2 dimensions, and thereby avoid the need for post-hoc dimensionality reduction, which itself can itself have a substantial impact on the obtained representation (\fig{S3-S4}). Ideally, we would wish to directly obtain 2D representations for the sequential models as well, but all attempts to train variants of the LSTM, ResNet, and Transformer models with 2D latent representations were unfruitful. This suggests that the additional complexity inherent in the sequential modeling of unaligned sequences places restrictions on how simple we can make the underlying latent representation (see discussion in Supplementary Material).

Continuous progress is being made in the area of sequential modeling and its use for protein representation learning \cite{shin2021protein,Heinzinger2019,rives2019biological, hawkins2021generating, rao2020transformer, elnaggar2020prottrans}. In particular, transformers, when scaled up to hundreds of millions of parameters, have been shown capable of recovering the covariances among sites in a protein \cite{rao2020transformer,vig2020bertology}.  When embedding the beta-lactamase sequences using these large pretrained transformer models, we indeed also see an improved separation of phyla (\fig{S5}). It remains an open question whether representations extracted from such large transformer models will eventually be able to capture more information than what can be extracted using a simple model and a high quality sequence alignment.

\emph{Implication: the scope of data (all proteins vs.\ single protein families), whether data is preprocessed into alignments, the model architecture, and potential post-hoc dimensionality reduction all have fundamental impact on the resulting representations, and the conclusions we can hope to draw from them. However, these contributions are often interdependent and difficult to disentangle in practice.}

\paragraph{Representation space topology carries relevant information.}


The star-like structure of the VAE representation in \fig\ref{fig:beta_lactamase}I, and the associated phyla color-coding strongly suggest that the topology of this particular representation space is related to the tree topology of the evolutionary history underlying the protein family \cite{Ding2019}. As an example of the potential and limits to representation interpretability, we will proceed with a more detailed analysis of this space.

\begin{figure}[b!] 
    \centering
    \includegraphics[width=0.48\textwidth]{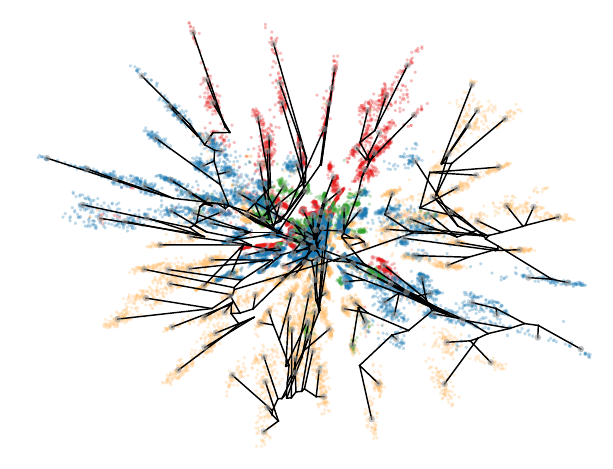}
    \caption{A phylogenetic tree encoded into the latent representation space. The representation and colors
      correspond to Fig.~\ref{fig:beta_lactamase}I. The internal nodes were determined using ancestral reconstruction after
      inferring a phylogenetic tree (branches encoded in black, leaf-nodes in gray).}
    \label{fig:points_tree}
\end{figure}

To explore the topological origin of the representation space, we estimate a phylogenetic tree of a subset of our input data ($n=200$), and encode the inner nodes of the tree to our latent space using a standard ancestral reconstruction method (see Methods). Although the fit is not perfect -- a few phyla are split and placed on opposite sides of the origin -- there is generally a good correspondence (\fig\ref{fig:points_tree}). We see that the reconstructed ancestors to a large extent span a meaningful tree, and it is thus clear that the representation topology in this case reflects relevant topological properties from the input space.  

\emph{Implication: Although neural networks are high capacity function estimators, we see empirically that topological constraints in input space are maintained in representation space. The latent manifold is thus meaningful and should be respected when relying on the representation for data interpretation.}

\begin{figure*}[t]
    \centering
    \begin{tabular}{p{0.20\textwidth} p{0.75\textwidth}}
    \vspace{5pt} \includegraphics[width=0.21\textwidth]{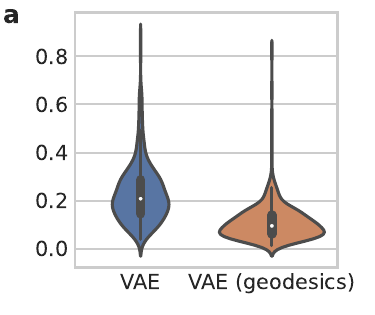} &
    \vspace{0pt} 
    \includegraphics[width=0.75\textwidth]{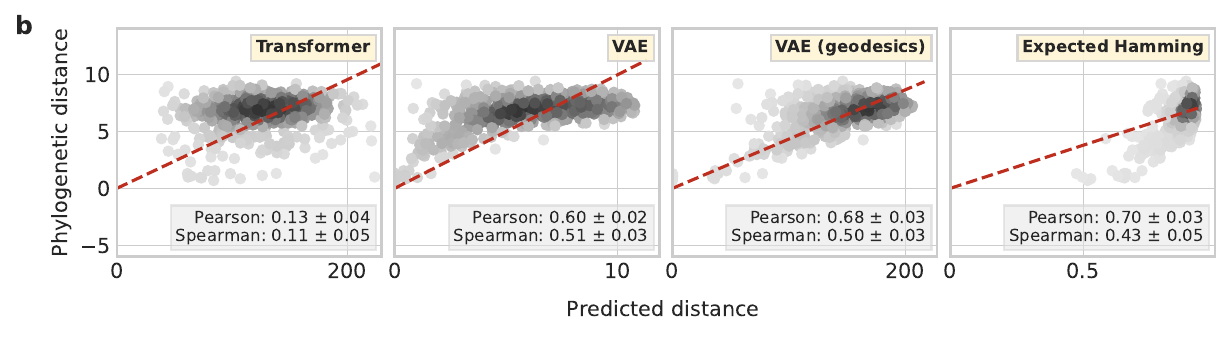}
\end{tabular}
    \caption{Geodesics provide more robust and meaningful distances in latent space. a) robustness of distances in latent space when calculated between the same data points embedded using models trained with different seeds. The plots show the distribution of standard deviations of normalized distances over five different models. b) Correlation between distances in latent space and phylogenetic distances, where standard deviations are calculated over 5 subsets of distances sampled with different seeds. Latent points were selected with probability proportional to their norm, to ensure a selection of distances covering the full range of latent space. The values for Transformer and VAE were calculated as Euclidean distances in their representation space (512 and 2 dimensional, respectively). }
    \label{fig:correlation}
\end{figure*}

\paragraph{Geometry gives robust representations.}
Perhaps the most exciting prospect of representation learning is the possibility of gaining new insights through interpretation and manipulation of the learned representation space. 
In NLP, the celebrated \texttt{word2vec} model \cite{Mikolov2013} demonstrated that simple arithmetic operations on representations yielded meaningful results, e.g. ``Paris - France + Italy = Rome'', and similar results are known from image analysis. The ability to perform such operations on proteins would have substantial impact on protein engineering and design, for instance making it possible to interpolate between biochemical properties or functional traits of a protein. What is required of our representations to support such interpolations?

To qualify the discussion, we note that standard arithmetic operations such as addition and subtraction rely on the assumption that the learned representation space is Euclidean. The star-like structure observed for the alignment-based VAE representation in \fig\ref{fig:points_tree} suggests that a Euclidean interpretation may be misleading: If we define similarities between pairs of points through the Euclidean distance between them, we implicitly assume straight-line interpolants that pass through uncharted territory in the representation space when moving between `branches' of the star-like structure. This does not seem fruitful.

Mathematically, the Euclidean interpretation is also problematic. In general, the latent variables of a generative model are not statistically identifiable, such that it is possible to deform the latent representation space without changing the estimated data density \cite{bishop:book, Hauberg2018}. The Euclidean topology is also known to cause difficulties when learning data manifolds with different topologies \cite{falorsi2018explorations, davidson2018hyperspherical}. With this in mind, the Euclidean assumption is difficult to justify beyond arguments of simplicity, as Euclidean arithmetic is not invariant to general deformations of the representation space. It has recently been pointed out that shortest paths (geodesics) and distances between representation pairs can be made identifiable even if the latent coordinates of the points themselves are not \cite{Arvanitidis2018, Hauberg2018}. The trick is to equip the learned representation with a Riemannian metric which ensures that distances are measured in data space along the estimated manifold. This result suggests that perhaps a Riemannian set of operations is more suitable for interacting with learned representations than the usual Euclidean arithmetic operators.

To investigate this hypothesis, we develop a suitable Riemannian metric, such that geodesic distances correspond to expected distances between one-hot encoded proteins, which are integrated along the manifold. The VAE defines a generative distribution $p(\X|\Z)$ that is governed by a neural network. Here $\Z$ is a latent variable, and $\X$ a one-hot encoded protein sequence. To define a notion of distance and shortest path we start from a curve $c$ in latent space, and ask what is its natural length? We parametrize the curve as $c: [0, 1] \rightarrow \mathcal{Z}$, where $\mathcal{Z}$ is the latent space, and write $c_t$ to denote the latent coordinates of the curve at time $t$. As the latent space can be arbitrarily deformed it is not sensible to measure the curve length directly in the latent space, and the classic geometric approach is to instead measure the curve length after a mapping to input space \cite{Hauberg2018}. For proteins, this amounts to measuring latent curve lengths in the one-hot encoded protein space. Shortest paths can then be found by minimizing curve length, and a natural distance between latent points is the length of this path.

An issue with this approach is that the VAE decoder is stochastic, such that the decoded curve is stochastic as well. To arrive at a practical solution, we recall that shortest paths are also curves of minimal \emph{energy} \cite{Hauberg2018} defined as

\vspace{-1em} 
\begin{align}
    \mathcal{E}(c) = \sum_{t=1}^{T-1} \norm{\X_{t+1} - \X_{t}}^2, 
\end{align}

\noindent
where $\X_t \sim p(\X | \Z = c_t)$ denote the protein sequence corresponding to latent coordinate $c_t$. Due to the stochastic decoder, the energy of a curve is a random variable. For continuous $\X$, recent work \cite{Arvanitidis2018} has shown promising results when defining shortest paths as curves with minimal \emph{expected} energy. In the Methods section we derive a similar approach for discrete one-hot encoded $\X$ and provide the details of the resulting optimization problem and its numerical solution.

To study the potential advantages of using geodesic over Euclidean distances, we analyze the robustness of our proposed distance. Since VAEs are not invariant to reparametrization we do not expect pairwise distances to be perfectly preserved between different initialization of the same model, but we hypothesize that the geodesics should provide greater robustness. We train the model 5 times with different seeds (see \fig{S9}) and calculate the same subset of pairwise distances. We normalize each set of pairwise distances by their mean and compute the distance standard deviation across trained models. When using normalized Euclidean distance we observe a mean standard deviation of $0.23$, while for normalized geodesics distances we obtain a value of $0.11$ (Fig.~\ref{fig:correlation}(a)). This significant difference indicates that geodesic distances are more robust to model retraining than their Euclidean counterparts.

\begin{figure*}[t!] 
    \centering
    \includegraphics[width=1.0\textwidth]{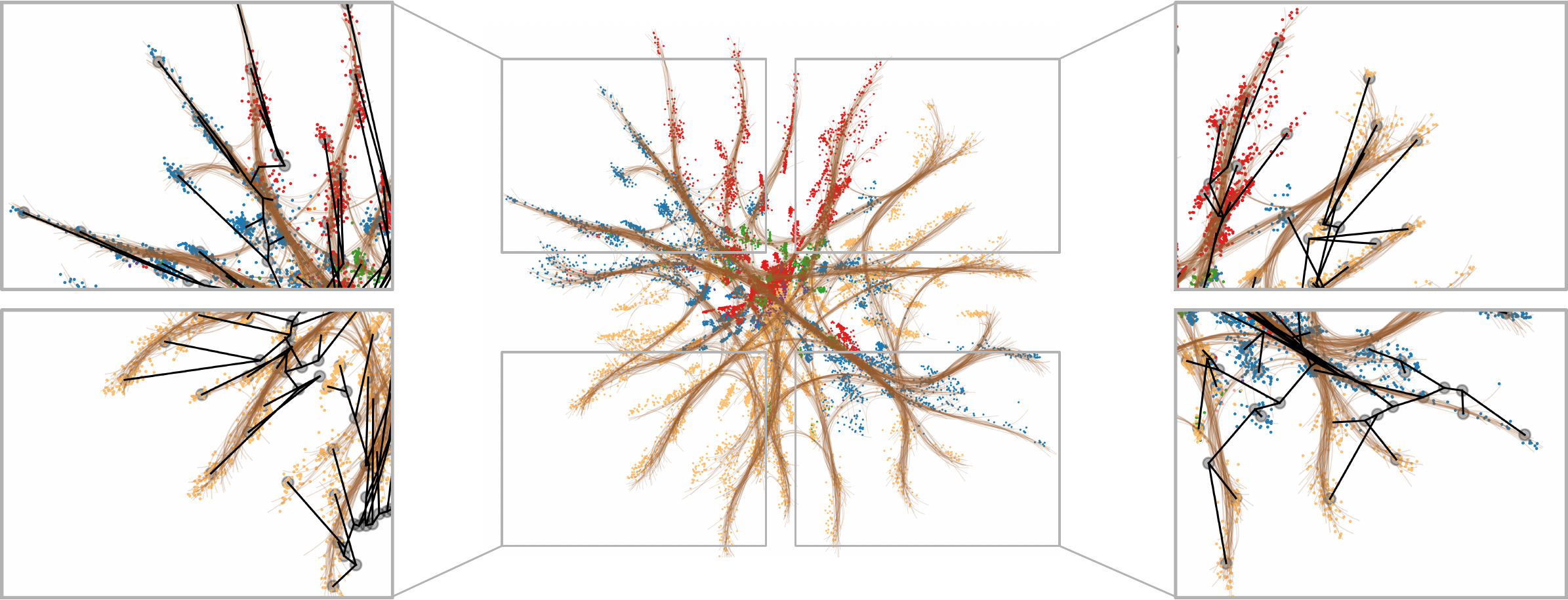}
    \caption{Shortest paths (geodesics) between representations of $\beta$-lactamase in a VAE. The Riemannian metric corresponds to measuring the expected distance between one-hot encoded proteins measured along the estimated manifold. The geodesics generally move along the star-shaped structure of the data similarly to the estimated phylogenetic tree, suggesting that the geodesics are well-suited for interpolating proteins.}
    \label{fig:many_geodesics}
\end{figure*}

\emph{Implication: distances and interpolation between points in representation space can be made robust by respecting the underlying geometry of the manifold.}

\paragraph{Geodesics give meaning to representations.}

To further investigate the usefulness of geodesics, we revisit the phylogenetic analysis of \fig\ref{fig:points_tree}, and consider how well distances in representation space correlate with the corresponding phylogenetic distances. The first two panels of Fig.~\ref{fig:correlation}(b) show the correlation between 500 subsampled Euclidean distances and phylogenetic distances in a Transformer and a VAE representation, respectively. We observe very little correlation in the Transformer representation, while the VAE fares somewhat better. The third panel of Fig.~\ref{fig:correlation}(b) shows the correlation between \emph{geodesic} distances and phylogenetic distances for the VAE. We observe that the geodesic distances significantly increases the linear correlation for particular short-to-medium distances. Finally, in the last panel, we include as a baseline the expected Hamming distance, i.e.~latent points decoded into their categorical distribution from which we draw 10 samples/sequences and calculate the average Hamming distance. We observe that the geodesics in latent space are a reasonable proxy for this expected distance in output space.

Visually, the correspondence is also striking (Fig.~\ref{fig:many_geodesics}).  Well optimized geodesics follow the manifold very closely, and to a large extent preserve the underlying tree structure. We see that the irregularities described before (e.g.\ the incorrect placement of the yellow subtree in the top right corner) are recognized by both the phylogenetic reconstruction and our geodesics, which is visually clear by the thick bundle of geodesics running diagonally to connect these regions. 

\emph{Implication: Analyzing geodesics distances instead of euclidean distances in representation space better reflects the underlying manifold allowing us to extract biological distances that are more meaningful.}

\paragraph{Data preprocessing affects the geometry}
We have established that the preprocessing of protein sequences into an alignment has a strong effect on the learned representation. But how do alignment quality and sequence selection biases affect the learned representations? To build alignments, it is common to start with a single \emph{query} sequence, and iterative search for sequences similar to this query. If the intent is to make statements only about this particular query sequence (e.g.\ predicting effects of variants relative to this protein) then a common practice is to remove columns in the alignment for which the query sequence has a gap. This query-centric bias is further enhanced by the fact that the search for relevant sequences occurs iteratively based on similarity, and is thus bound to have greater sequence coverage for sequences close to the query. These effects would suggest that representations learned from query-centric alignments might be better descriptions of sequences close to the query.

\begin{figure}[t!]
    \centering
    \includegraphics[width=0.49\textwidth]{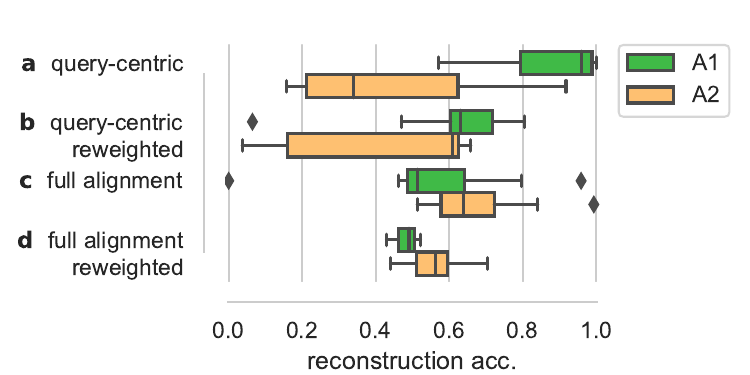}
    \caption{The effect of alignment preprocessing on the ability of representations to reliably decode back to protein sequences. Box plots (median, upper/lower quartiles, 1.5 inter-quartile range) show the distribution of reconstruction accuracies across the 14 class A1 and 13 class A2 sequences. Query-centric denotes an alignment where columns in the alignment have been removed if they contain a gap in the query sequence of interest. Reweighted refers to the standard practice of reweighting protein sequences based on similarity to other sequences. All four cases contain the same protein sequences. A1 and A2 are subclasses of beta-lactamase. A2 sequences have substantially worse representations when alignments are focused on a query from the A1 class.}
    \label{fig:different_alignments}
\end{figure}

\begin{figure*}[t]
    \centering
    \includegraphics[width=1.0\textwidth]{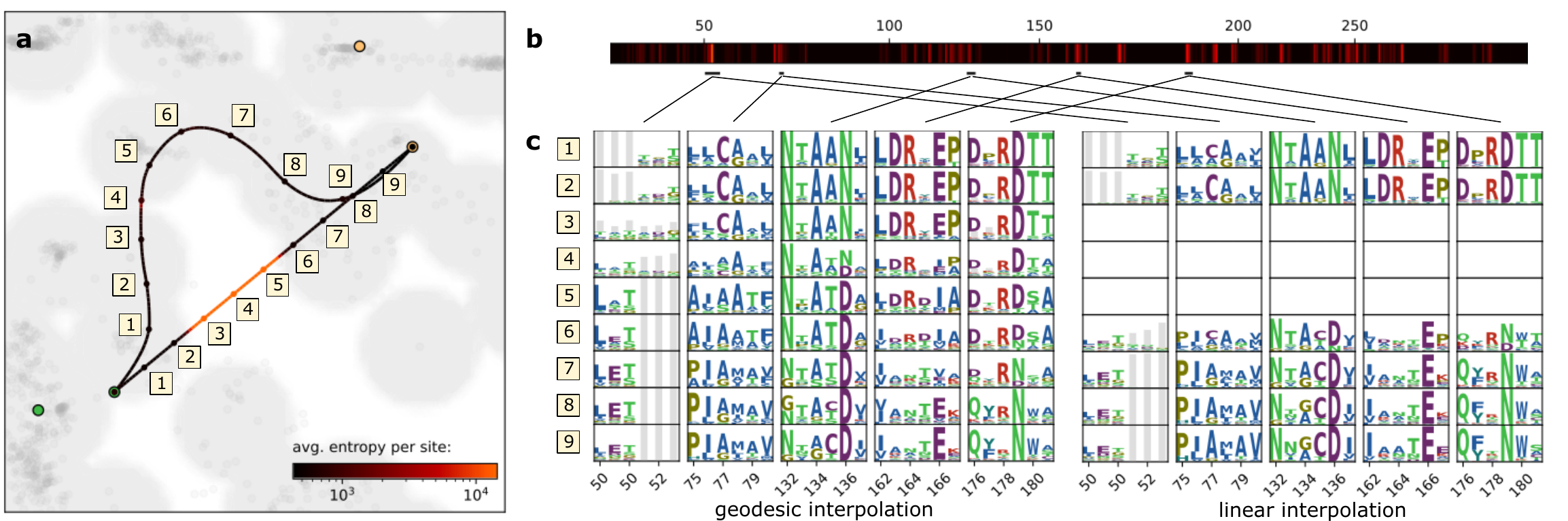}
    \caption{Interpolation between two protein sequences. a) The latent space corresponding to the bottom-right representation in Fig.~\ref{fig:different_alignments}, where we have selected a sequence from $\beta$-lactamase subclass A1 and subclass A2, and consider two interpolation paths between the proteins: a direct, linear interpolation and one following the geodesic. The curves are color coded by the entropy of the amino acid output distribution along the path. b) The Kullback-Leibler (relative entropy) of the target sequence relative to the source for each position in the alignment. Regions with large difference (i.e. high relative entropy) are highlighted in red. c) The output distributions corresponding to several of the high-entropy regions, for the different points along the interpolant. Distributions are encoded using the weblogo standard, where the height of the letter-column at each position encodes how peaked the distribution is. }
    \label{fig:protein_interpolation2}
\end{figure*}

To test this hypothesis, we look at a more narrow subset of the $\beta$-lactamase family, covering only the class A $\beta$-lactamases. This subset was included as part of the DeepSequence paper \cite{Riesselman2018} and will serve as our representative example of a query-centric alignment.  The class A $\beta$-lactamases consist of two subclasses, A1 and A2, which are known to display consistent differences in multiple regions of the protein. The query sequence in this case is the TEM from Escherichia coli, which belongs to  subclass A1. Following earlier characterization of the differences between the subclasses, we consider a set of representative sequences from each of the subclasses, and probe how they are mapped to representation space (Class A1: TEM-1, SHV-1, PSE-1, RTG-2, CumA, OXY-1, KLUA-1, CTX-M-1, NMCA, SME-1, KPC-2, GES-1, BEL-1, BPS-1. Class A2: PER-1, CEF-1, VEB-1, TLA-2, CIA-1, CGA-1, CME-1, CSP-1, SPU-1, TLA-1, CblA, CfxA, CepA). When training a representation model on the original alignment (\fig\ref{fig:different_alignments}a), we indeed see that the ability to reconstruct (decode) meaningful sequences from representation values differs dramatically between the A1 and A2 classes.

It is common practice to weigh input sequences in alignments by their density in sequence space, which compensates for the sampling bias mentioned above \cite{Ekeberg2013}. While this is known to improve the quality of the model for the variant effect prediction \cite{Riesselman2018}, it only partially compensates for the underlying bias between the classes in our case (\fig\ref{fig:different_alignments}b). If we instead retrieve full length sequences for all proteins, redo the alignment using standard software (Clustal Omega \cite{clustalo}), and maintain the full alignment length, we see that the differences between the classes becomes much smaller (\fig\ref{fig:different_alignments}c-d). The reason is straightforward: as the distance from the query sequence increases, larger parts of a protein will occur within the regions corresponding to gaps in the query sequence. If such columns are removed, we discard more information about the distant sequences, and therefore see larger uncertainty (i.e.\ entropy) for the decoder of such latent values. Note that these differences in representation quality are not immediately clear through visual inspection alone (\fig{S7}).

\emph{Implication: Alignment-based representations depend critically on the nature of the multiple sequence alignment. In particular, training on query-centric alignments results in representations that primarily describe
sequence variation around a single query sequence. In general, density based reweighting of sequences should be used to counter selection bias.}

\paragraph{Geodesics provide more meaningful interpolation.}

The output distributions obtained by decoding from representation space provide interpretable insights into the nature of the representation. We illustrate this by constructing an interpolant along the geodesic from a subclass A1 member to a subclass A2 member (\fig\ref{fig:protein_interpolation2}a).  We calculate the entropy of the output distribution (summed over all sequence positions) along the interpolant and observe that there is a clear transition with elevated entropy around point 5 (highlighted in red). To investigate which regions of the protein are affected, we calculate the Kullback-Leibler divergence between the output distributions of the end points (\fig\ref{fig:protein_interpolation2}b). Zooming in on these particular regions (\fig\ref{fig:protein_interpolation2}c, left), and following them along the interpolant, we see that the representation naturally captures transitions between amino acid preferences at different sites. Most of these correspond to sites already identified in prior literature, for instance disappearance of the cysteine at position 77, the switch between N $\rightarrow$ D at position 136, and D $\rightarrow$ N at position 179 
\cite{philippon2016structure}. We also see an example where a region in one class aligns to a gap in the other (position 50-52). The linear interpolation (\fig\ref{fig:protein_interpolation2}c, right), has similar statistics at the endpoints, but displays an almost trivial interpolation trajectory, which effectively interpolates linearly between the probability levels of the output classes at the end points (note for instance the minor preference for cysteine in the A2 region at position~77).

\emph{Implication: Geodesics provide natural interpolants between points in representation space, avoiding high entropy regions, and thereby providing interpolated values that are better supported by data.}

\section*{Discussion}

Learned representations of protein sequences can substantially improve systems for making biological predictions, and may also help to reveal previously uncovered biological information. In this paper, we have illuminated parts of the answer to the titular question of what constitutes a meaningful representation of proteins. One of the conclusions is that the question itself does not have a single general answer, and must always be qualified with a specification of the purpose of the representation. A representation that is suitable for making predictions may not be optimal for a human investigator to better understand the underlying biology, and vice versa. The enticing idea of a single protein representation for all tasks thus seems unworkable in practice.

\subsection*{Designing purposeful representations} 

Designing a representation for a given task requires reflection over which biological properties we wish the representation to encapsulate.  Different biological aspects of a protein will place different demands on the representations, but it is not straightforward to enforce specific properties in a representation. We can, however, steer the representation learning by 1) picking appropriate model architectures, 2) preprocessing the data, 3) choosing suitable objective functions, and 4) placing prior distributions on parts of the model. We discuss each of these in turn.

\textbf{Informed network architectures} can be difficult to construct as the usual neural network `building blocks' are fairly elementary mathematical functions that are not immediately linked to high-level biological information. Nonetheless, our discussion of length-invariant sequence representations is a simple example of how one might inform the model architecture of the biology of the task. It is generally acknowledged that global protein properties are not linearly related to local properties. It is therefore not surprising when we show that the model performance significantly improves when we allow the model to learn such a nonlinear relationship instead of relying on the common linear average of local representations. It would be interesting to push this idea beyond the Resnet architecture that we explored here, in particular in combination with the recent large scale transformer-based language models. We speculate that while similar `low-hanging fruit' may remain in currently applied network architectures, they are limited, and more advanced tools are needed to encode biological information into network architectures. The internal representations in attention-based architectures have been shown to recover known physical interactions between proteins \cite{rao2020transformer, vig2020bertology}, opening the door to the incorporation of prior information about known physical interactions in a protein. Recent work on permutation and rotation invariance/equivariance in neural networks \cite{cohen2018intertwiners,weiler20183d} hold promise, though they have yet to be explored exhaustively in representation learning. 

\textbf{Data preprocessing} and feature engineering is frowned upon in contemporary `end-to-end' representation learning, but it remains an important part of model design. In particular, preprocessing using the vast selection of existing tools from computational biology is a valuable way to encode existing biological knowledge into the representation. 
We saw a significant improvement in the representation capabilities of unsupervised models when trained on aligned protein sequences, as this injects prior knowledge about comparable sequence positions in a set of sequences. While recent work is increasingly working towards techniques for learning such signals directly from data \cite{shin2021protein,rao2020transformer,vig2020bertology}, it remains unclear if the advantages provided by multiple alignments can be fully encapsulated by these methods. Other preprocessing techniques, such as the reweighing of sequences, are currently also dependent on having aligned sequences. These examples suggests that if we move too fast towards `end-to-end' learning, we risk throwing the baby out with the bathwater, by discarding years of experience endowed in existing tools.

\textbf{Relevant objective functions} are paramount to any learning task. Although representation learning is typically conducted using a reconstruction loss, we demonstrate that optimal representations according to this objective are generally sub-optimal for any specific transfer-learned task. 
This suggests that hyper-parameters of representations should be chosen based on downstream task-specific performance, rather than reconstruction performance on a hold-out set. This is, however, a delicate process, as optimizing the \emph{parameters} of the representation model on the downstream task is associated with a high risk of overfitting. We anticipate that principled techniques for combining reconstruction objectives on the large unsupervised data sets with task specific objectives in a semi-supervised learning setting will provide substantial benefits in this area \cite{Min2019}.

\textbf{Informative priors} can impose softer preferences than those encoded by hard architecture constraints.
The Gaussian prior in VAEs is such an example, though its preference is not guided by biological information, which appears to be a missed opportunity. In the studies of $\beta$-lactamase, we, and others \cite{Riesselman2018,Ding2019}, observe a representation structure that resembles the phylogenetic tree spanned by the evolution of the protein family. Recent hyperbolic priors \cite{mathieu2019continuous} that are designed to emphasize hierarchies in data may help to more clearly bring forward such evolutionary structure. Since we observe that the latent representation better reflects biology when endowed with a suitable Riemannian metric, it may be valuable to use corresponding geometric priors \cite{kalatzis:icml:2020}.\looseness=-1

\subsection*{Analyzing representations appropriately}

Even with the most valiant efforts to incorporate prior knowledge into our representations, they must still be interpreted with great care. We highlight the particular example of distances in representation space, and emphasize that the seemingly natural Euclidean distances are misleading. The non-linearity of encoders and decoders in modern machine learning methods means that representation spaces are generally non-Euclidean. We have demonstrated that by bringing the expected distance from the observation space into the representation space in the form of a Riemannian metric, we obtain geodesic distances that correlate significantly better with phylogenetic distances than what can be attained through the usual Euclidean view. This is an exciting result as the Riemannian view comes with a set of natural operators akin to addition and subtraction, such that the representation can be engaged with operationally. We expect this to be valuable for e.g.\ protein engineering, since it gives an operational way to combine representations from different proteins. 

In this study, we employed our geometric analysis only on the latent space of a variational autoencoder, which is well-suited due to its smooth mapping from a fixed dimensional latent space to a fixed dimensional output space. Expanding beyond single protein families is hindered by the
fact that we cannot decode from an aggregated global representation in a sequential language model. A natural question is whether Bottleneck strategies like the one we propose could make such analysis possible. If so, it would present new possibilities for defining meaningful distances between remote homologues in latent space \cite{morton2020protein}, and potentially allow for improved transfer of GO/EC annotations between proteins.

Finally, the geometric analysis comes with several implications that are relevant beyond proteins. It suggests that the commonly applied visualizations where latent representations are plotted as points on a Euclidean screen may be highly misleading. We therefore see a need for visualization techniques that faithfully reflect the geometry of the representations. The analysis also indicates that downstream prediction tasks may gain from leveraging the geometry, although standard neural network architectures do not yet have such capabilities.


\small{
\section*{Methods}

\paragraph{Variational autoencoders.}
A variational autoencoder assumes that data $\X$ is generated from some (unknown) latent factors $\Z$ though the process $p_\theta (\X| \Z)$. The latent variables $\Z$ can be viewed as the compressed representation of $\X$. Latent space models try to model the joint distribution of $\X$ and $\Z$ as $p_\theta(\X, \Z) = p_\theta(\Z) p_\theta(\X | \Z)$. The generating process can then be viewed as a two step procedure: first a latent variable $\Z$ is sampled from the prior and then data $\X$ is sampled from the conditional $p_\theta(\X | \Z)$ (often called the decoder). Since $\X$ is discrete by nature, $p_\theta(\X | \Z)$ is modeled as a Categorical distribution $p_\theta(\X | \Z) \sim Cat(C, l_\theta(\Z))$ with $C$ classes and $l_\theta(Z)$ being the log-probabilities for each class. To make the model flexible enough to capture higher order amino acid interactions, we model $l_\theta(Z)$ as a neural network. Even though data $\X$ is discrete, we use continuous latent variables $\Z \sim N(0,1)$.

\paragraph{Construction of entropy network.}
To ensure that our VAE decodes to high uncertainty in regions of low data density, we construct an explicit network architecture with this property. That is, the network $p_{\theta}(\X | \Z)$ should be certain about its output in regions where we have observed data, and uncertain in regions where we have not. This has been shown to be important to get well-behaved Riemannian metrics \cite{Hauberg2018, Tosi:UAI:2014}. In a standard VAE with posterior modeled as a normal distribution $\mathcal{N}(\mu_\theta(\Z), \sigma^2_\theta(\Z))$, this amounts to constructing a variance network $\sigma^2_\theta(\Z)$ that increases away from data \cite{Arvanitidis2018, Detlefsen2019}. However, no prior work has been done on discrete distributions, such as the Categorical distribution $C(\mu_\theta(\Z))$ that we are working with. In this model we do not have a clear division of the average output (mean) and uncertainty (variance), so we control the uncertainty through the entropy of the distribution. We remind that for a categorical distribution, the entropy is
\begin{linenomath}$$ H(\X|\Z)=\sum_{i=1}^C p_\theta(\X|\Z)_i \cdot \log p_\theta(\X|\Z)_i. $$\end{linenomath}
The most uncertain case corresponds to when $H(\X|\Z)$ is largest i.e.\ when $p(\X|\Z)_i=1/C$ for $i=1,...,C$. Thus, we want to construct a network $p_\theta(\X|\Z)$ that assigns equal probability to all classes when we are away from data, but is still flexible when we are close to data. Taking inspiration from \cite{Detlefsen2019} we construct a function $\alpha=T(z)$, that maps distance in latent space to the zero-one domain ($T: [0, \inf) \mapsto [0,1]$). $T$ is a trainable network of the model, with the functional form $T(z)=\texttt{sigmoid}\big(\frac{-6.9077\beta \cdot V(z)}{\beta}\big)$ with $V(z)=\min_{j=\{1,..,K\}} \norm{z-c_j}_2^2$, where $c_j$ are trainable cluster centers (initialized using $k$-means). This function essentially estimates how close a latent point $z$ is to the data manifold, returning 1 if we are close and 0 when far away. Here $K$ indicates the number of cluster centers (hyperparameter) and $\beta$ is a overall scaling (trainable, constrained to the positive domain). With this network we can ensure a well-calibrated entropy by picking

$$ p_\theta(\X|\Z)_i = \alpha \cdot p_\theta(\X|\Z)_i + (1-\alpha) \cdot \mathbb{L}, $$
where $\mathbb{L}=\frac{1}{C}$. For points far away from data, we have $\alpha=0$ and return $\mathbb{L}$ regardless of category (class), giving maximal entropy. When near the data, we have $\alpha=1$ and the entropy is determined by the trained decoder $p_\theta(\X|\Z)_i$.

Figure~\ref{fig:entropy} shows the difference in entropy of the likelihood between a standard VAE (left) and a VAE equipped with our developed entropy network (right). The standard VAEs produce arbitrary entropy, and is often more confident in its predictions far away from the data. Our network increases entropy as we move away from data.

\begin{figure}[t!]
    \centering
    \includegraphics[width=0.45\textwidth]{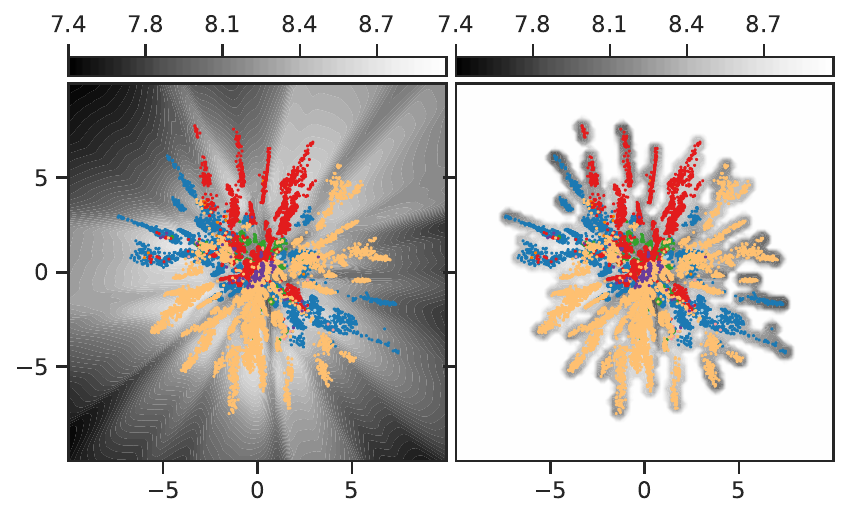}
    \caption{Construction of the entropy network for our geodesic calculations. Left: latent representations of $\beta$-lactamase with the background color denoting the 
    entropy of the output posterior for a standard VAE. Right: as left but using a VAE equipped with our developed entropy network.}
    \label{fig:entropy}
\end{figure}

\paragraph{Distance in sequence space.} To calculate geodesic distances we first need to define geodesics over the random manifold defined by $p(\X|\Z)$. These geodesics are curves $c$ that minimize expected \emph{energy} \cite{Hauberg2018} defined as
\begin{linenomath}\begin{equation}
    \bar{\mathcal{E}}(c)
      = \mathbb{E}[\mathcal{E}(c)]
      = \int_0^1 \mathbb{E}\left[ \norm{\partial_t \X_t }^2 \right] \: \mathrm{d}t,
    \label{eq:energy_measure}
\end{equation} \end{linenomath}
where $\X_t \sim p(\X | \Z = c_t)$ is the decoding of a latent point $c_t$ along the curve $c$. This energy requires a  meaningful (squared) norm in data space. We remind here that protein sequence data $x, y$ is embedded into a one-hot space i.e.
\begin{linenomath}$$x,y\in\left\{ \begin{bmatrix} 1 \\ 0 \\ \vdots \\ 0 \end{bmatrix},\begin{bmatrix} 0 \\ 1 \\ \vdots \\ 0 \end{bmatrix},\begin{bmatrix} 0 \\ 0 \\ \vdots \\ 1 \end{bmatrix}\right\}, $$\end{linenomath}
where we assume that $p(x_d=1)=a_d$, $p(y_d=1)=b_d$ for $d=1,...,C$. It can easily be shown that the squared norm between two such one-hot vectors can either be 0 or 2:
\begin{linenomath}$$\Delta^2 = \norm{x-y}^2 = \{0,2\}.$$\end{linenomath}
The probability of these two events are given as
\begin{linenomath}\begin{align*}
    P(\Delta^2=0) &= P(x=y) \\
    &=P(x_1=y_1) + P(x_2=y_2) + ... + P(x_D=y_D) \\
    &=\sum_{d=1}^C a_d b_d, \\
    P(\Delta^2=2)&=1-P(\Delta^2 =0)
    =1-\sum_{d=1}^C a_d b_d.
\end{align*}\end{linenomath}
The expected squared distance is then given by
\begin{linenomath}\begin{align*}
    \mathop{\mathbb{E}}(\Delta^2) &= \int_{\{0,2\}} \Delta^2 \cdot P(\Delta^2) d\Delta^2 \\
    &=0 \cdot P(\Delta^2=0) + 2 \cdot P(\Delta^2=2) \\
    &=2\left(1- \sum_{d=1}^C a_d b_d \right),
\end{align*}\end{linenomath}
Extending this measure to two sequences of length $L$ is then
\begin{linenomath}\begin{equation}
    \mathop{\mathbb{E}}(\Delta^2) = \sum_{l=1}^L 2\left(1- \sum_{d=1}^C a_{l,d} b_{l,d}\right).
    \label{eq:distance_measure}
\end{equation}\end{linenomath}  
The energy of a curve, can then be evaluated by integrating this sequence measure \eqref{eq:distance_measure}
along the given curve,
\begin{linenomath}\begin{equation}
  \bar{\mathcal{E}}(c) \approx 2 \sum_{i=1}^{N-1}\!\! \sum_{l=1}^L\! \left(1 \!-\! \sum_{d=1}^C p(c_i)_{l,d}\  p(c_{i+1})_{l,d} \right) \Delta t,
  \label{eq:disc_energy}
\end{equation}\end{linenomath}
where $\Delta t = \norm{c_{i+1}-c_{i}}_2$.
Geodesics can then be found by minimizing this energy \eqref{eq:disc_energy} with respect to the unknown curve $c$. For an optimal curve $c$, its length is given by $\sqrt{\bar{\mathcal{E}}(c)}$.

\paragraph{Optimizing geodesics.}
In principal, the geodesics could be found by direct minimization of the expected energy. However, empirically we observed that this strategy was prone to diverge, since the optimization landscape is very flat near the initial starting point. We therefore instead discretize the entropy landscape into a 2D grid, and form a graph based on this. In this graph each node will be a point in the grid, which is connected to its eight nearest neighbors, with the edge weight being the distance weighted with the entropy. Then, using Dijkstra's algorithm \cite{Dijkstra59anote} we can rapidly find a robust initialization of each geodesic. To obtain the final geodesic curve we fit a cubic spline \cite{meir_1968} to the discretized curve found by Dijkstra's algorithm, and afterwards do 10 gradient steps over the spline coefficients with respect to the curve energy \eqref{eq:disc_energy} to refine the solution. 

\paragraph{Phylogeny and ancestral reconstruction.}
The $n=200$ points used for the ancestral reconstruction were chosen as latent embeddings from the training set that were closest to the trainable cluster centers $\{c_i\}_{i=1}^n$ found during the estimation of the entropy network. We used FastTree2 \cite{Price2010} with standard settings for estimation of phylogenetic trees and subsequently applied the codeml program \cite{adachi1996molphy} from the PAML package for ancestral reconstruction of the internal nodes of the tree.
}

\paragraph{Data availability.}
All data used in this manuscript originates from publicly available databases. The specific sequence data for pre-training and data for the different protein task can be found online: \url{https://github.com/songlab-cal/tape}. Data for the $\beta$-lactamase familie can be found here: \url{https://pfam.xfam.org/family/PF00144}. Preprocessed data is available through the scripts provided in our code repository.

\paragraph{Code availability.}
The source code for the paper is freely available online under an open source licence: \url{https://github.com/MachineLearningLifeScience/meaningful-protein-representations}.

\paragraph{Competing interests.}
The authors declare no competing interests.



\printbibliography

\paragraph{Acknowledgements.}
This work was funded in part by the Novo Nordisk Foundation through the Center for Basic Machine Learning Research in Life Science (NNF20OC0062606). It also received funding from the European Research Council (ERC) under the European Union’s Horizon 2020 research and innovation programme (757360). NSD and SH were supported in part by a research grant (15334) from VILLUM FONDEN. WB was supported by a project grant from the Novo Nordisk Foundation (NNF18OC0052719). We thank Ole Winther, Jesper Ferkinghoff-Borg, and Jesper Salomon for feedback on earlier versions of this manuscript. Finally, we gratefully acknowledge the support of NVIDIA Corporation with the donation of GPU hardware used for this research.

\paragraph{Author contributions. }
NSD, SH and WB jointly conceived and designed the study. NSD and WB conducted the experiments. All authors contributed to the writing of the paper.

\typeout{get arXiv to do 4 passes: Label(s) may have changed. Rerun}

\end{document}


\renewcommand{\thefigure}{S\arabic{figure}}
\renewcommand{\thetable}{S\arabic{table}}

\maketitle

\section*{Experimental details}

\paragraph{Datasets.} In the transfer-learning experiments we use 31 million protein sequences extracted from the Pfam database \cite{pfam}, following the procedure described for the TAPE benchmark set \cite{Rao2019}. 
We use data for remote homology detection from \cite{Hou2018}, fluorescence landscape prediction from \cite{Sarkisyan2016} and for stability landscape prediction from \cite{Rocklin2017}, all spanning multiple protein families. See Table~\ref{tap:data_size} for the specific dataset sizes. \\

For the experiments regarding the analysis of reconstruction error as a measure of downstream performance (\ref{fig:latent_size}), we use the UniLanguage dataset \cite{Armenteros2020}. UniLanguage consists of samples from the UniProt \cite{uniprot} database, where splits were constructed to minimize the overlap between families. 

For the study of latent space structure on single protein families we consider the $\beta$-lactamase sequences extracted from Pfam \cite{pfam}, family PF00144, where we also obtain a sequence alignment. For the study of subclasses of beta-lactamase, we use the beta-lactamase alignment provided in the DeepSequence study \cite{Riesselman2018}, which is processed in different ways as described in the main paper. The processing scripts are provided as part of the source code repository associated with our manuscript.

\begin{table}[h!]
\begin{tabular}{l|l|l|l|}
Task              & Train      & Valid & Test        \\ \hline
Language Mod. & 32,207,059 & N/A   & 44,314       \\
Unilanguage \cite{Armenteros2020} & 607,737 & 98,907 & 295,161 \\
Remote Homol.   & 12,312     & 736   & 718         \\
Fluorescence      & 21,446     & 5,362 & 27,217      \\
Stability         & 53,679     & 2,447 & 12,839   
\end{tabular}
\caption{Data set sizes used for the prediction tasks (i.e. the transfer-learning setting).}
\label{tap:data_size}
\end{table}

\paragraph{Predictive tasks.}
In the transfer-learning experiments, the Transformer and Resnet based models were pre-trained using a masked token prediction task \cite{Devlin2019} where 15\% of the amino acids in a sequence are masked out and the task is to predict the identity of the masked amino acids from the non-masked. The LSTM model were trained using next-token prediction, where the task is to predict the next amino acid in a sequence given the amino acids processed until now. Lastly, the autoencoder (bottleneck) models were trained with standard reconstruction tasks. Details for the three downstream tasks are listed below:
\begin{enumerate}
 \item \underline{Fluorescence}: An input protein sequence $\textbf{s}$ is mapped to a label $y \in \mathbb{R}$ corresponding to the log-fluorescence intensity of $\textbf{s}$, that expresses a models ability to distinguish between similar sequences. The models are optimized using the mean squared loss and performance is measured using Spearman correlation.
 \item \underline{Stability}: An input protein sequence $\textbf{s}$ is mapped to a label $y \in \mathbb{R}$ corresponding to the most extreme value for which the protein keeps its fold. The models are optimized using the mean squared loss and performance is measured using Spearman correlation.
 \item \underline{Remote homology}: An input protein sequence $\textbf{s}$ is mapped to a label $y \in \{1,...,1195\}$, where each class correspond to a specific protein fold. The models are optimized using categorical cross entropy and performance is measured using accuracy. 
\end{enumerate}

\section*{Network Architectures}

\subsection*{Resnet, LSTM, Transformer}
The Resnet, LSTM and Transformer architectures used in the first experiments are all directly taken from TAPE \cite{Rao2019}. The Transformer consist of 12-layers with a hidden size of 512 and 8 attention heads, which leads to a 38M-parameter model. The architectures of the two other models were chosen such that the total number of parameters match that of the Transformer. In this case, the Resnet consist of 35 layers each with 256 filters, a dilation rate of 2 and a kernel size of 9. The LSTM has 3 bidirectional layers each with 1024 hidden units. If not stated otherwise, we use an attention based aggregation function for combining the local representations into a single global representation.

\subsection*{Bottleneck AutoEncoder}

For the Bottleneck Resnet autoencoder we used an encoder-decoder architecture where both the encoder and decoder were modeled using resnet blocks. In contrast to the three models above, the bottleneck autoencoder is not a sequential model and requires a fixed size input. All sequences were therefore padded with zeros to the same length (3000). For the encoder we use 30 residual blocks with pooling along the sequence dimension every 5 layer. For the decoder we inverse the process and again use 30 residual blocks, this time with upsampling along the sequence dimension every 5 layer. Between the encoder and decoder we had two fully connected layers that respectively downsampled and upsampled from the global latent space. The AutoEncoder has approximately twice the number of parameters as the sequential models during pre-training, but was designed to have the same number of parameters when used for transfer-learning as the decoder is disabled in this setting.

\subsection*{VAE}

The architecture of the VAE is a simplified version of that used in the DeepSequence paper \cite{Riesselman2018}, using an encoder with two fully connected hidden layers (both with 1500 nodes) with ReLU activations, and a decoder with two fully connected hidden layers (100 and 500 nodes) also with ReLU activations.

\paragraph{Training details.} We followed the training protocol from \cite{Rao2019} for pretraining on Pfam and training of the task specific models. Pre-training was performed on four NVIDIA TITAN V GPUs for 1 week. Hyperparameters were set as follows:
\begin{itemize}
    \item Adam optimizer was used with default settings for momentum.
    \item Learning rate: initialized to $10^{-3}$, adjusted using a linear warm-up scheduler.
    \item 10\% dropout rate.
    \item Batch size was dynamically set during training to the largest possible based on model architecture and sequence length.
\end{itemize}
Task-specific training was performed using the same set of GPUs and hyperparameters, but training was stopped early when no increase in validation performance was observed. If not stated otherwise, we always complete pre-training before task-specific training to get the best possible performing model.

For the training on $\beta$-lactamase model we deploy nearly the same training strategy as with the Pfam family, however with extra steps to prevent overfitting, which is more likely on a single family than the full corpus of proteins. In particular we use early stopping monitored on the validation loss with a patience of 10 epochs to ensure that we do not use highly overfitted models.

The VAEs were trained using the same optimizer settings, but with a fixed learning rate, no dropout and using a fixed batch size of 16.

\section*{Additional results}

\begin{figure}[b!] 
    \centering
    \includegraphics[width=0.4\textwidth]{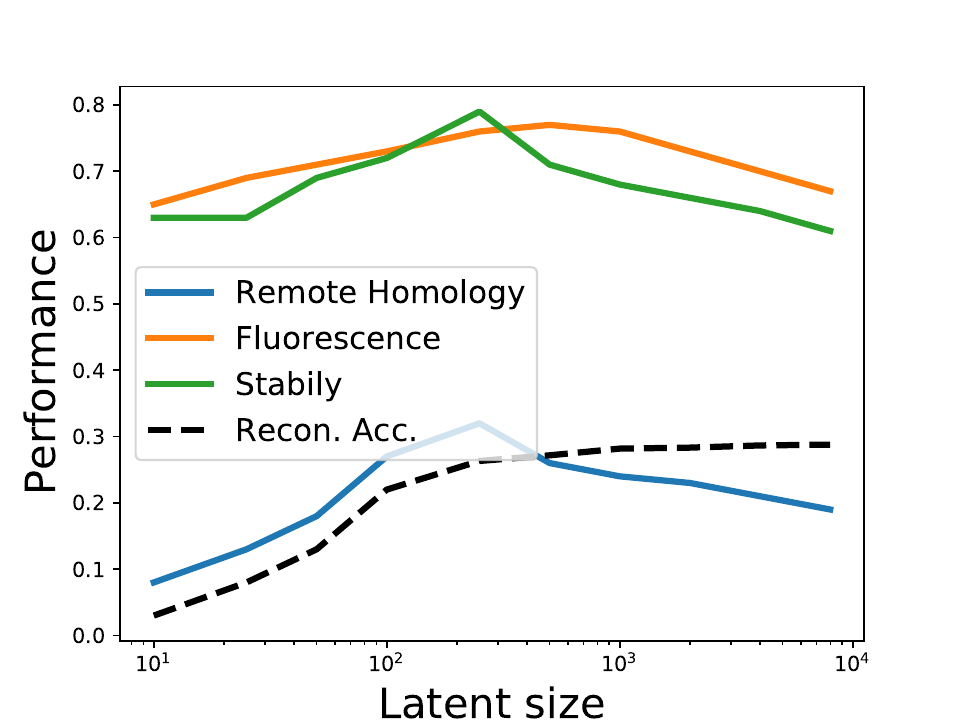}
    \caption{Reconstruction and downstream performance as a function of representation size. Although reconstruction accuracy consistently improves for increasing representation size, the performances on the individual tasks deteriorate for large representations. Performance refers to Spearman correlation (stability, fluorescence) or accuracy (homology, reconstruction). }
    \label{fig:latent_size}
\end{figure}

\subsection*{Reconstruction accuracy is not a good proxy for  downstream performance}
\begin{figure} 
    \centering
    \includegraphics[width=0.35\textwidth]{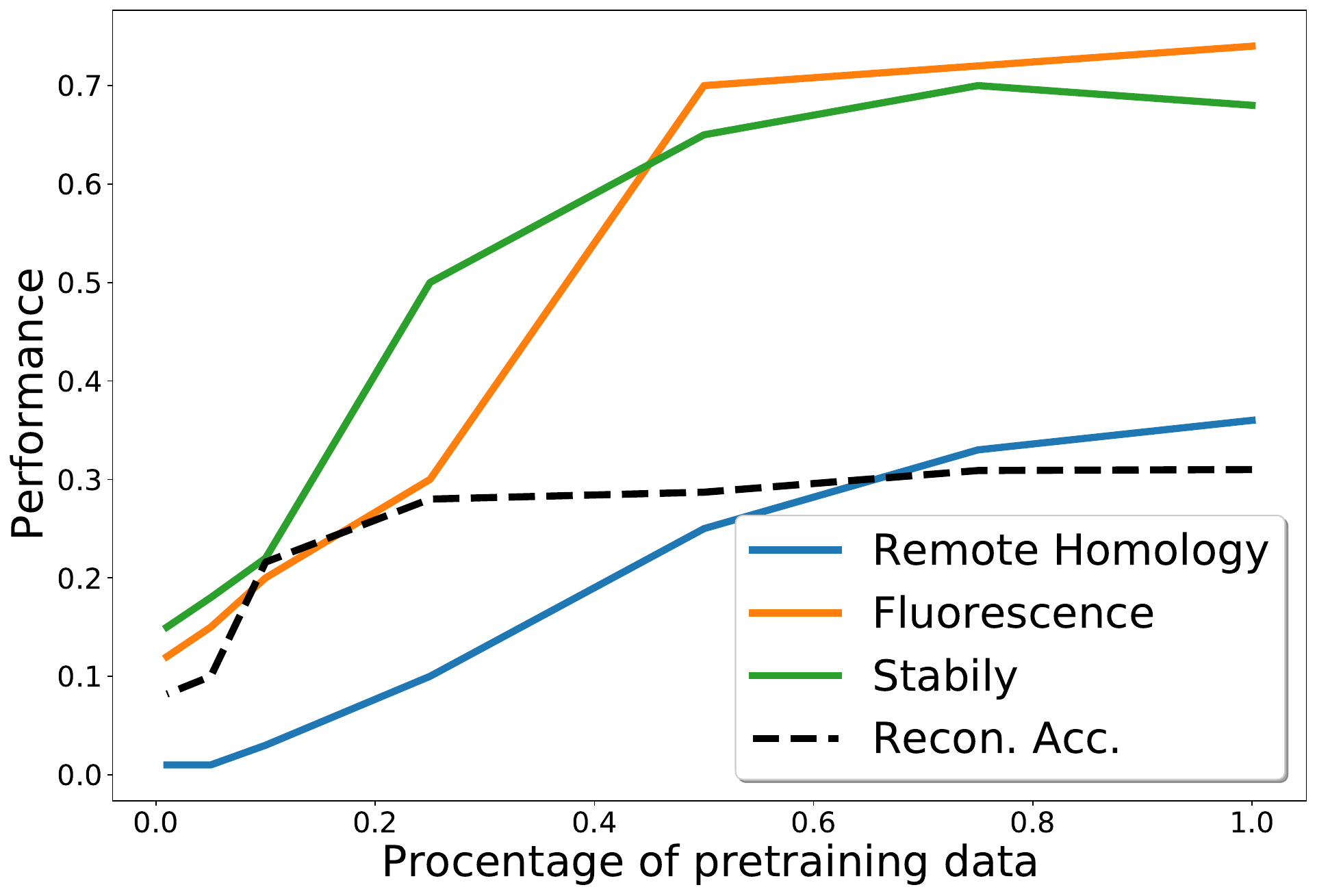}
    \caption{Reconstruction and downstream performance as a function of amount of data used during pre-training (in \%). Performance refers to Spearman correlation (stability, fluorescence) or accuracy (homology, reconstruction).}
    \label{fig:data_size}
\end{figure}

As stated in the main paper, we observe that reconstruction accuracy may be a poor proxy for the quality of the representation itself, as it does not directly correlate with the downstream performance metrics (\fig\ref{fig:latent_size}). To further investigate the phenomenon, we re-trained a number of embedding models, where we gradually lowered the amount of pre-training data available to the model (\fig\ref{fig:data_size}). We again observe a discrepancy between the reconstruction performance and the downstream performance metrics, with the reconstruction accuracy flattening out after seeing only 30\% of the data, whereas the downstream tasks all increase with more pre-training data. This confirms the discussion from the main paper that the reconstruction accuracy of a model is a poor proxy for the how well the representation will perform on downstream tasks.

\subsection*{Impact of modeling choices on representation}

In the main paper, we discuss how learned representations are affected by data preprocessing, choice of modeling architecture and post-hoc dimensionality reduction. It would be convenient to avoid the post-hoc dimensionality reduction step altogether. This is possible for the VAE, by simply setting the latent dimension to 2, but turned out to be infeasible for the sequential models. Even in the case of the bottleneck ResNet, which in many ways is similar to the VAE, it was not possible to train models with a two dimensional representation, which suggests that the low dimensional bottleneck is incompatible with the requirements for the expressivity of the encoder/decoder in these more complex sequential models.

\begin{figure}[t!]
    \centering
    \includegraphics[width=0.46\textwidth]{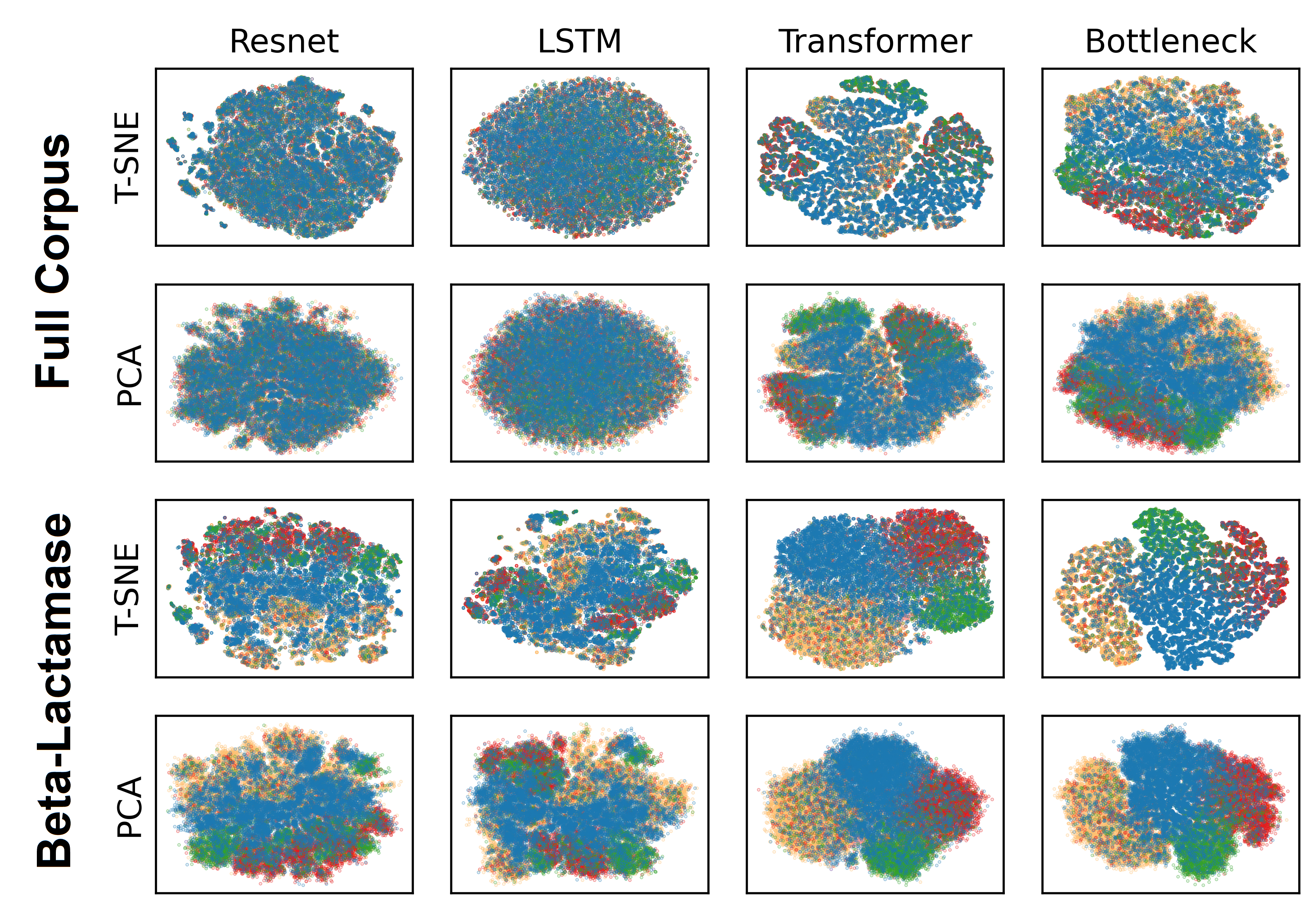}
    \caption{Impact of the choice of dimensionality reduction on the representation manifolds. This figure corresponds to the \fig{2} in the main paper, but includes dimensionality reduction with both PCA and t-SNE.}
    \label{fig:dim_reduce1}
\end{figure}

\begin{figure}[t!] 
    \centering
    \includegraphics[width=0.45\textwidth]{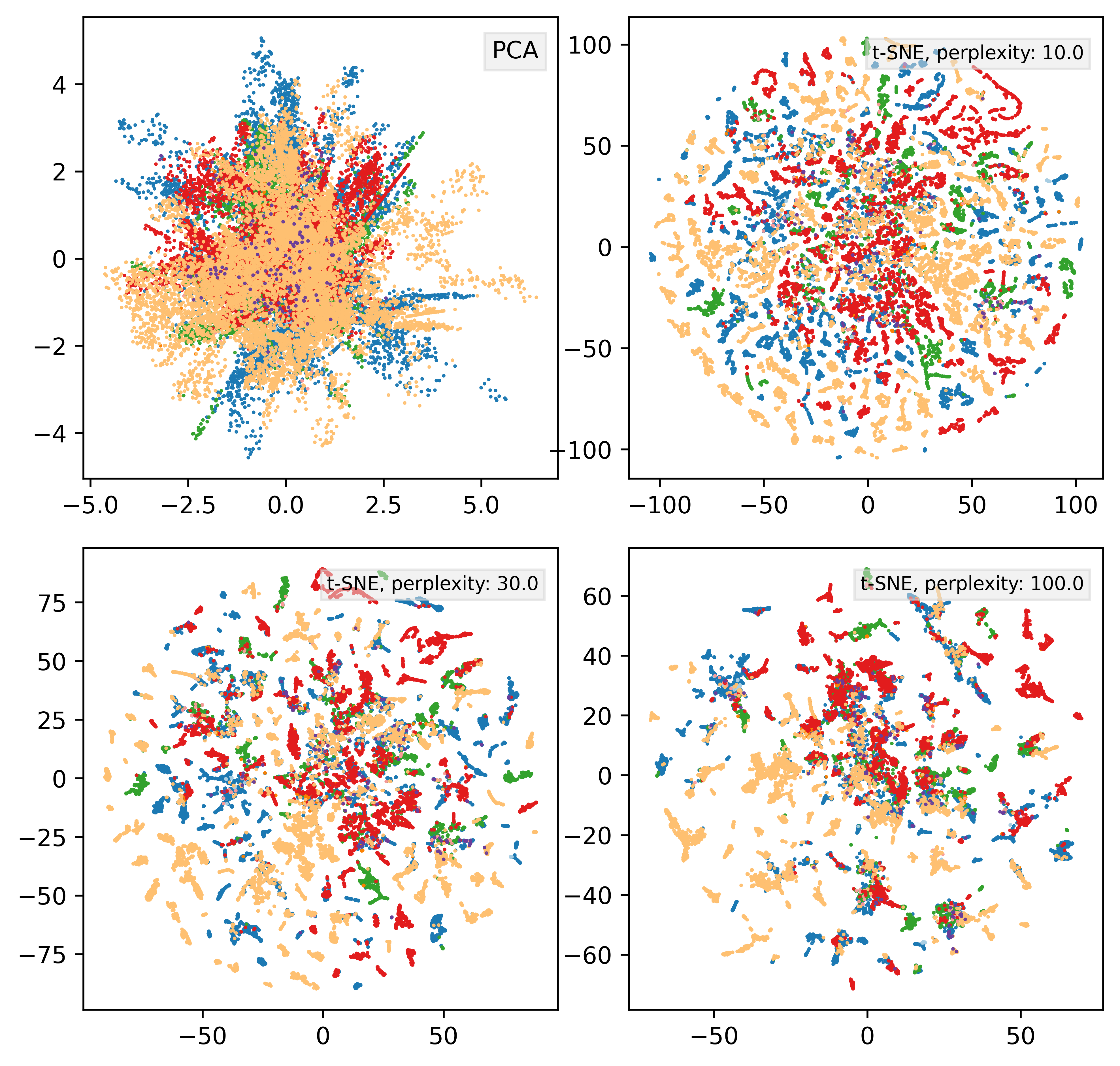}
    \caption{Dimensionality reduction on the latent space of a VAE. In contrast to the two dimensional latent space employed in the main paper, we here train a model with 30 latent dimensions. The two plots show the same latent space reduced by either PCA or t-SNE, the latter with different choices of the perplexity parameter.}
    \label{fig:dim_reduce2}
\end{figure}

To test the impact of the dimensionality reduction step on the representation, we conducted dimensionality reduction for all sequential models with both PCA and t-SNE (\fig\ref{fig:dim_reduce1}).
Although the t-SNE representations seem to have slightly better separation of phyla locally, the overall patterns produced by the two dimensionality reduction schemes are quite robust, in particular in the most specific representations in the bottom right. We also investigated the impact of dimensionality reduction for the VAE, when trained with a higher dimensional latent space of 30, and applying t-SNE or PCA to reduce it to two dimensions (\fig\ref{fig:dim_reduce2}). In this case, the PCA seems to preserve the star-like structure to some extent, while t-SNE organizes the clusters in a completely different topology, presumably due to the distributional assumptions underlying the t-SNE method.

\begin{figure*}[!t]
    \centering
    \includegraphics[width=0.7\textwidth]{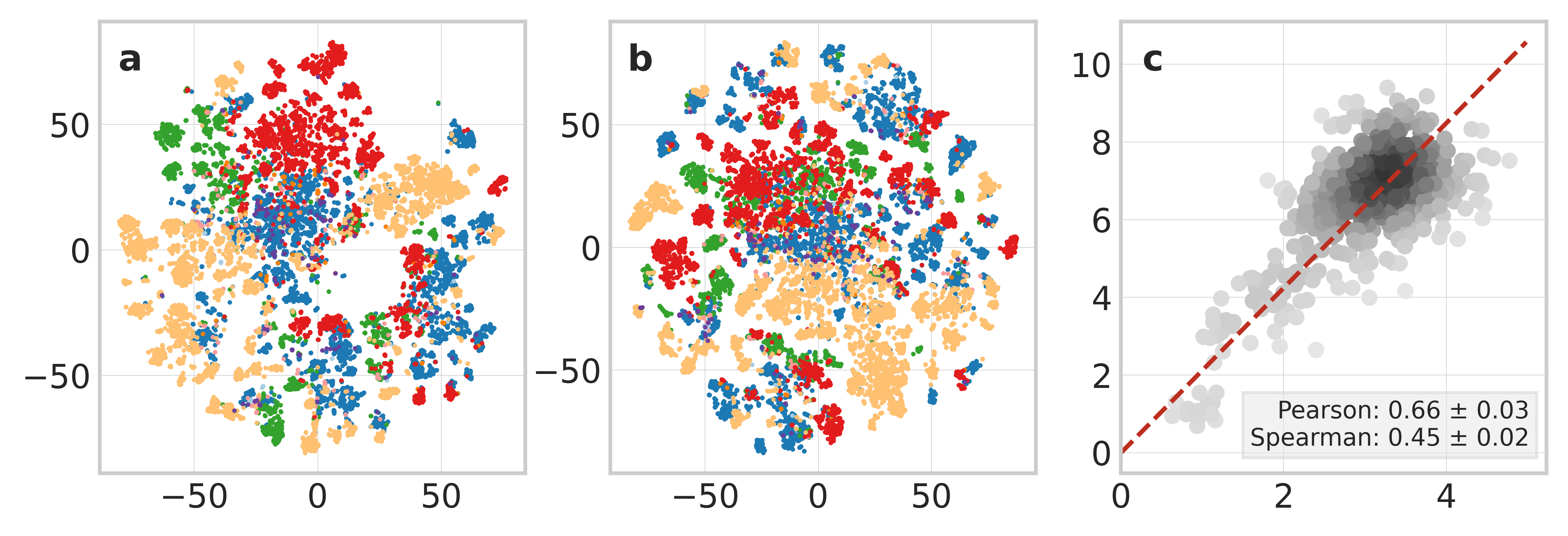}
    \caption{Representations extracted from a state-of-the-art large transformer model, ESM-1b \cite{rives2019biological}. The embeddings were calculated in two different ways: a) extracting sequences from the beta-lactamase multiple sequence alignment used in main paper by removing gaps, b) retrieving the full-length sequences corresponding to the proteins in the alignment. The two approaches lead to comparable representations. Compared to the shallow transformer presented in the main paper, we see that large scale language models capture similar information as obtained from a multiple sequence alignment - at least in terms of correlation to phylogenetic distances (c).}
    \label{fig:esm1b}
\end{figure*}

One aspect of model architecture that was not explored in the main paper was the size of the model, in terms of the number of parameters. The universal, "full-corpus", models explored in \fig2 in the main paper were all trained on the same dataset, with a similar number of parameters. In \fig\ref{fig:esm1b} we show the behavior of a state-of-the-art language model, ESM-1b, which has substantially more parameters, and is trained on a much larger dataset \cite{rives2019biological}. We see that this larger-scale transformer model displays improved separation capabilities compared to the other models in the top row in \fig2, and that the correlation to phylogenetic distances becomes comparable to that of models trained on the specific family.

\begin{figure}
    \centering
    \includegraphics[width=0.48\textwidth]{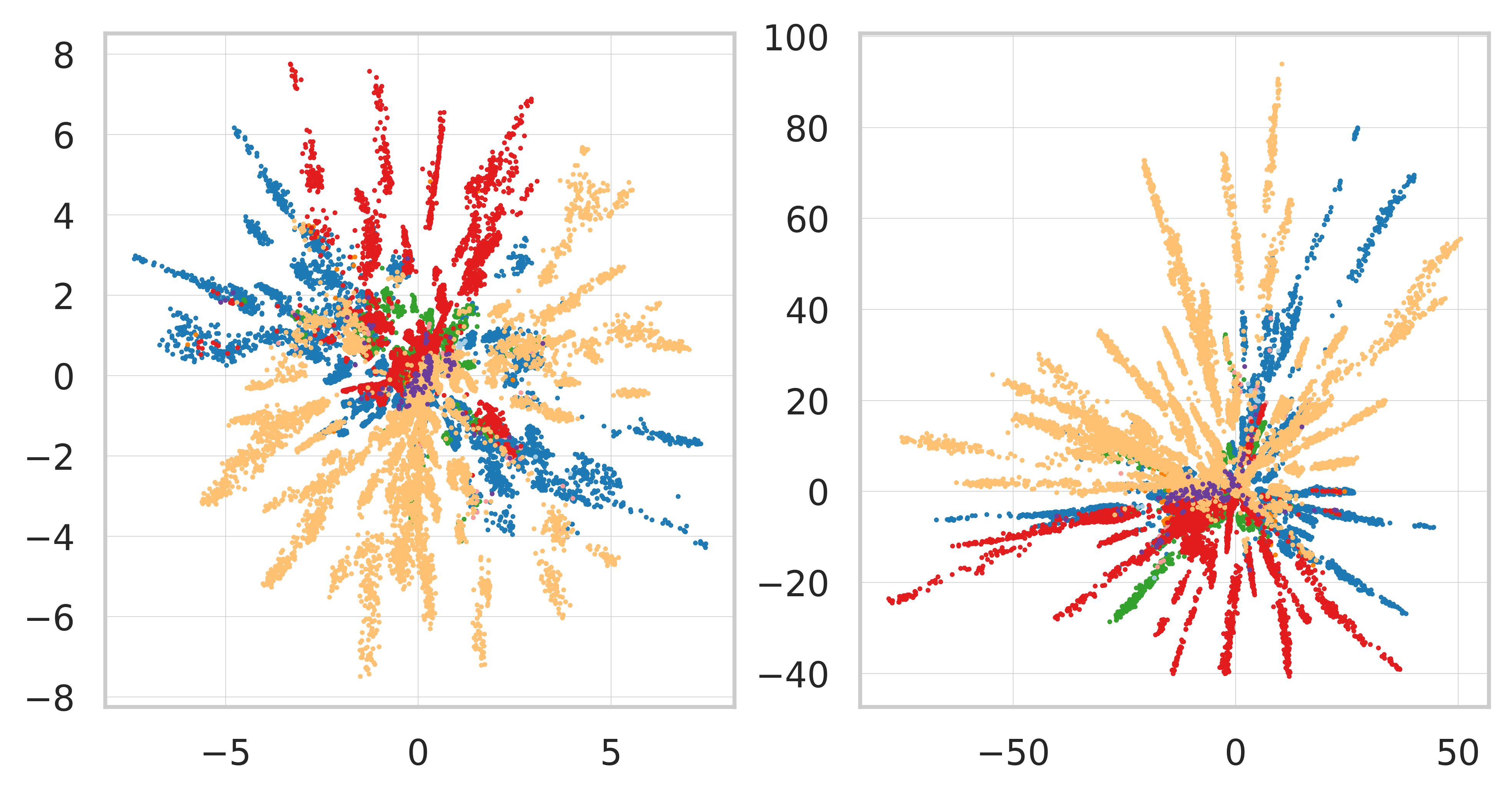}
    \caption{Illustration of the difference in obtained representation between a variational autoencoder (left) and a standard autoencoder (right). One of the main differences is that the $\mathcal{N}(0,1)$ prior in the variational autoencoder sets the scale of the latent space. }
    \label{fig:vae-vs-ae}
\end{figure}

While some architecture choices clearly have an impact on the resulting representation, other choices seem to be of little consequence. For instance, we observe that a standard (non-variational) autoencoder produces very similar representations as a variational autoencoder, when trained on the same data set. The inductive bias difference between these two models thus seems to be negligible, apart from the regularizing $\mathcal{N}(0,1)$ prior which sets an overall scale in representation space (\fig\ref{fig:vae-vs-ae}).

\begin{figure}[t!]
    \centering
    \includegraphics[width=0.49\textwidth]{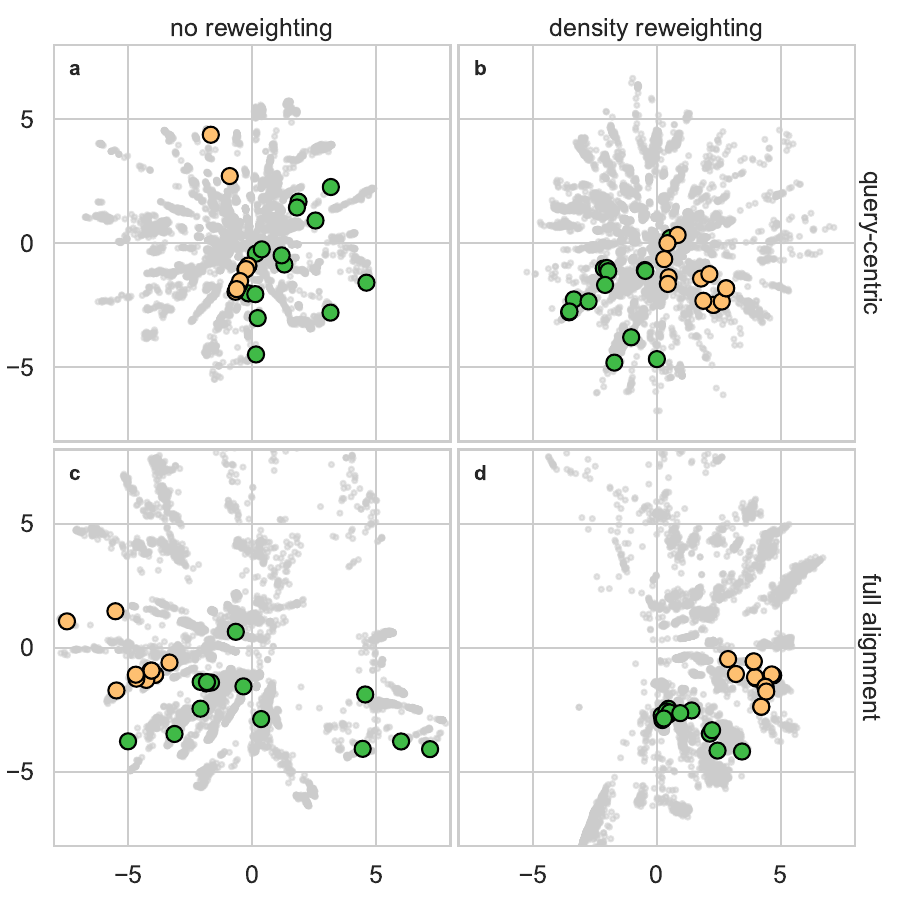}
    \caption{The effect of alignment preprocessing on the learned representation. Top row: query-centric alignment where columns are removed if they contain a gap in the query sequence. Bottom row: standard alignment of the same sequences. Left/Right column: whether the sequences are reweighted during training of the model. The green dots correspond to proteins belonging to the same subclass as the query (A1). The yellow dots belong to subclass A2, which is more distant to the query protein.}
    \label{fig:different_alignments}
\end{figure}

\begin{figure*}[t!]
    \centering
    \includegraphics[width=0.725\textwidth]{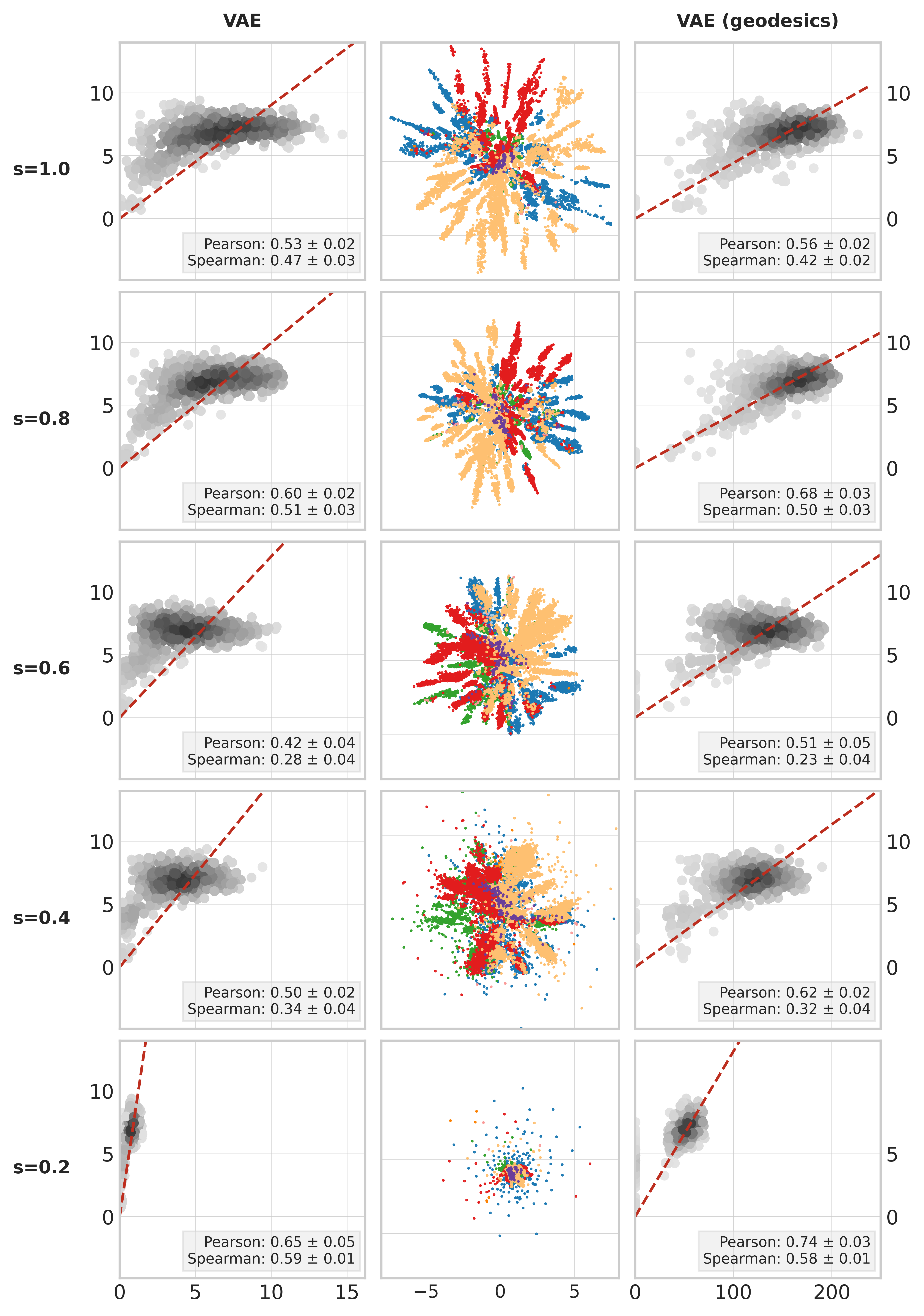}
    \caption{The effect of sequence reweighting on the obtained representation. The rows display representations obtained with VAEs trained with different degrees of density reweighting of the input sequences. The $s$ parameter denotes the maximum hamming distance used as a cutoff to define a neighborhood of similar sequences \cite{Ekeberg2013}. The $s=0.8$ setting corresponds to what is referred to as \emph{density reweighting} in the main text and \fig\ref{fig:different_alignments}.}
    \label{fig:reweighted}
\end{figure*}

\subsection*{Effect of alignment preprocessing}
In the main text, we discuss how reweighting of input data and column removal in the alignment can lead to representations with different degrees of selection bias towards a particular query sequence (\fig6). Visualizations of the four settings discussed in the main paper are displayed in \fig\ref{fig:different_alignments}. 

\subsection*{Degree of sequence reweighting} The VAE in \fig3 in the main paper was trained without conducting sequence reweighting. For completeness, we here include results from training on a dataset where the input sequences are reweighted based on input density, for different choices of the hamming cutoff distance $s$, which defines how large a neighborhood is considered when estimating the density \cite{Ekeberg2013}. As expected, we see a better balanced representation of the different phyla when reweighting, most pronounced in the added weight on the green Firmicutes subtree (\fig\ref{fig:reweighted}) and the even lower populated phyla that are difficult to discern in the original VAE. The trends in the correlations to phylogenetic distances are similar to those reported in the main paper, again demonstrating a benefit of employing geodesics rather than euclidean distances. 

\subsection*{Impact of initialization on the representation} 
\begin{figure}[h!]
    \centering
    \includegraphics[width=0.48\textwidth]{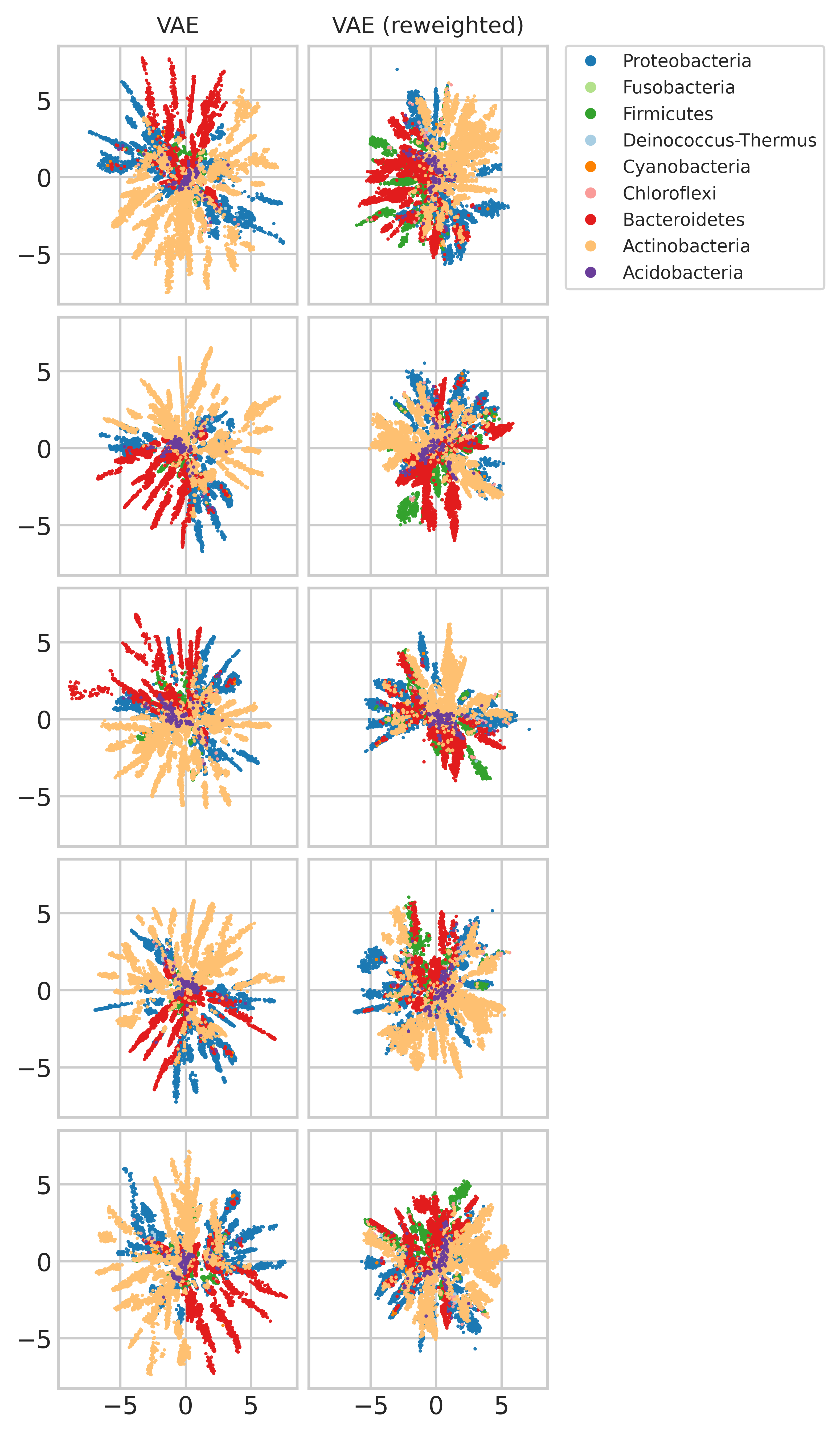}
    \caption{Different initializations of the same model. Left column: trained on raw data; right column: trained on density reweighted data ($s=0.8$). The top left plot corresponds to the model used in the main paper.}
    \label{fig:inits}
\end{figure}

Although it is commonly accepted that initialization of a neural network has some impact on the resulting models, the ultimate behavior and performance of a model is often fairly robust to different initializations. It is important to stress that this is not the case for learned representations, which can change dramatically depending on the initialization, partly due to the many symmetries in parameter space. As an example, in \fig\ref{fig:inits} we show the representations of the $\beta$-lactamase protein family for 4 different initial seeds. While they all follow the overall tree structure, we see clear variations in the organization of the individual branches of the tree, and we especially observe that the orientation in latent space is arbitrary. 
This further supports the idea that a Euclidean interpretation of the latent space can be misleading. \fig4a in the main paper quantifies this effect in terms of the robustness of distances calculations across different training instances.

\printbibliography

\typeout{get arXiv to do 4 passes: Label(s) may have changed. Rerun}